\documentclass[journal]{IEEEtran}

\newif\iffull
\fulltrue

\usepackage{algorithmic, algorithm}
\usepackage[english]{babel}
\usepackage{amssymb}
\usepackage{amsmath,amsfonts}
\usepackage{array}
\usepackage{booktabs}
\usepackage[caption=false,font=normalsize,labelfont=sf,textfont=sf]{subfig}
\usepackage{textcomp}
\usepackage{stfloats}
\usepackage{url}
\usepackage{verbatim}
\usepackage{mathtools}
\usepackage{mathrsfs}
\usepackage{xcolor}
\usepackage{cite}
\usepackage{graphicx}
\usepackage{amsthm}
\usepackage{bm}
\usepackage{comment}
\usepackage[normalem]{ulem}
\usepackage{enumitem}
\usepackage[hidelinks]{hyperref}
\usepackage{thm-restate}
\usepackage{breqn}
\usepackage{soul}

\theoremstyle{definition}
\newtheorem{theorem}{Theorem}
\newtheorem{lemma}{Lemma}
\newtheorem{claim}{Claim}
\newtheorem{definition}{Definition}
\newtheorem{corollary}{Corollary}
\newtheorem{problem}{Problem}
\newtheorem{construction}{Construction}
\newtheorem{example}{Example}
\newtheorem{encoder}{Encoder}

\newtheorem{remark}{Remark}

\newcommand{\bfh}{{\boldsymbol h}}

\newcommand{\bfv}{{\boldsymbol v}}

\newcommand{\bfX}{{\mathbf X}}
\newcommand{\bfY}{{\mathbf Y}}
\newcommand{\bfZ}{{\mathbf Z}}

\newcommand{\cA}{\mathcal{A}}
\newcommand{\cB}{\mathcal{B}}
\newcommand{\cC}{\mathcal{C}}
\newcommand{\cD}{\mathcal{D}}
\newcommand{\cE}{\mathcal{E}}

\newcommand{\cH}{\mathcal{H}}
\newcommand{\cI}{\mathcal{I}}
\newcommand{\cJ}{\mathcal{J}}

\newcommand{\cL}{\mathcal{L}}

\newcommand{\cX}{\mathcal{X}}

\renewcommand{\Bbb}{\mathbb}

\newcommand{\F}{{\Bbb F}}

\newcommand{\Mod}[1]{\ (\mathrm{mod}\ #1)}

\usepackage{stackengine}
\def\delequal{\mathrel{\ensurestackMath{\stackon[1pt]{=}{\scriptstyle\Delta}}}}

\def\BibTeX{{\rm B\kern-.05em{\sc i\kern-.025em b}\kern-.08em
    T\kern-.1667em\lower.7ex\hbox{E}\kern-.125emX}}
\usepackage{balance}

\title{\textbf{Tail-Erasure-Correcting Codes}}
\author{Boaz Moav,~\IEEEmembership{Student,~IEEE,} Ryan Gabrys,~\IEEEmembership{Member,~IEEE,} Eitan Yaakobi,~\IEEEmembership{Senior Member,~IEEE}
\thanks{B. Moav and E. Yaakobi are with the Henry and Marilyn Taub Faculty of Computer Science, Technion - Israel Institute of Technology, Haifa 3200003, Israel 
(e-mail: \texttt{\{boazmoav,yaakobi\}@cs.technion.ac.il}). 
R. Gabrys is with University of California, San Diego, California, USA   
(e-mail: \texttt{rgabrys@ucsd.edu}).}
\thanks{%
The research was funded by the European Union (ERC, DNAStorage, 865630). Views and opinions expressed are however those of the authors only and do not necessarily reflect those of the European Union or the European Research Council Executive Agency. Neither the European Union nor the granting authority can be held responsible for them. This work was also supported in part by NSF Grant CCF2212437.}}

\IEEEoverridecommandlockouts

\begin{document}
\maketitle

\begin{abstract}
The increasing demand for data storage has prompted the exploration of new techniques, with molecular data storage being a promising alternative. 
In this work, we develop coding schemes for a new storage paradigm that can be represented as a collection of two-dimensional arrays. 
Motivated by error patterns observed in recent prototype architectures, our study focuses on correcting erasures in the last few symbols of each row, and also correcting arbitrary deletions across rows.
We present code constructions and explicit encoders and decoders that are shown to be nearly optimal in many scenarios. 
We show that the new coding schemes are capable of effectively mitigating these errors, making these emerging storage platforms potentially promising solutions.
\end{abstract}

\begin{IEEEkeywords}
Coding theory, DNA data storage, deletions, tail-erasure, RT-metric.
\end{IEEEkeywords}

\section{Introduction}
\label{sec:introduction}
\IEEEPARstart{T}{he} DNA Data Storage market is rapidly growing and is expected to approach \$2 billion (USD) in valuation by 2028 \cite{DNA23}. This rapid growth is driven by several promising technologies that depart from existing storage practices.
Traditional approaches to DNA storage, including the pioneering works of \cite{CGK12,GBCDLS13}, typically represent data using individual base pairs of DNA (i.e., adenine (A) and thymine (T), cytosine (C) and guanine (G)).

More recently, companies and researchers have moved away from this paradigm, and in an effort to drive down the cost of synthesis and overcome some of the practical limitations of reading with traditional DNA sequencers, they have begun to investigate using collections of DNA molecules, sometimes referred to as cassettes, in order to encode logical '0's and `1's, such as in the Iridia system~\cite{Iridia19}. These DNA cassettes are  sequentially chained together and stored within an atomic unit, which is referred to as a \textit{nano-memory cell (NMC)} within a larger storage architecture.

In order to accommodate these new storage paradigms while also keeping the broader (traditional) storage technology stack intact, these memory cells are logically organized into larger collections of cells analogous to sectors in a hard drive. This organization is beneficial in several ways. First, it allows one to parallelize the read/write process enabling faster access times. In addition, this logical organization of NMCs can leverage addressing schemes that are typically employed in the context of traditional magnetic media storage  thus reducing the potential cost or impact for transition.  

\begin{figure}
\centering
\includegraphics[width=0.9\linewidth]{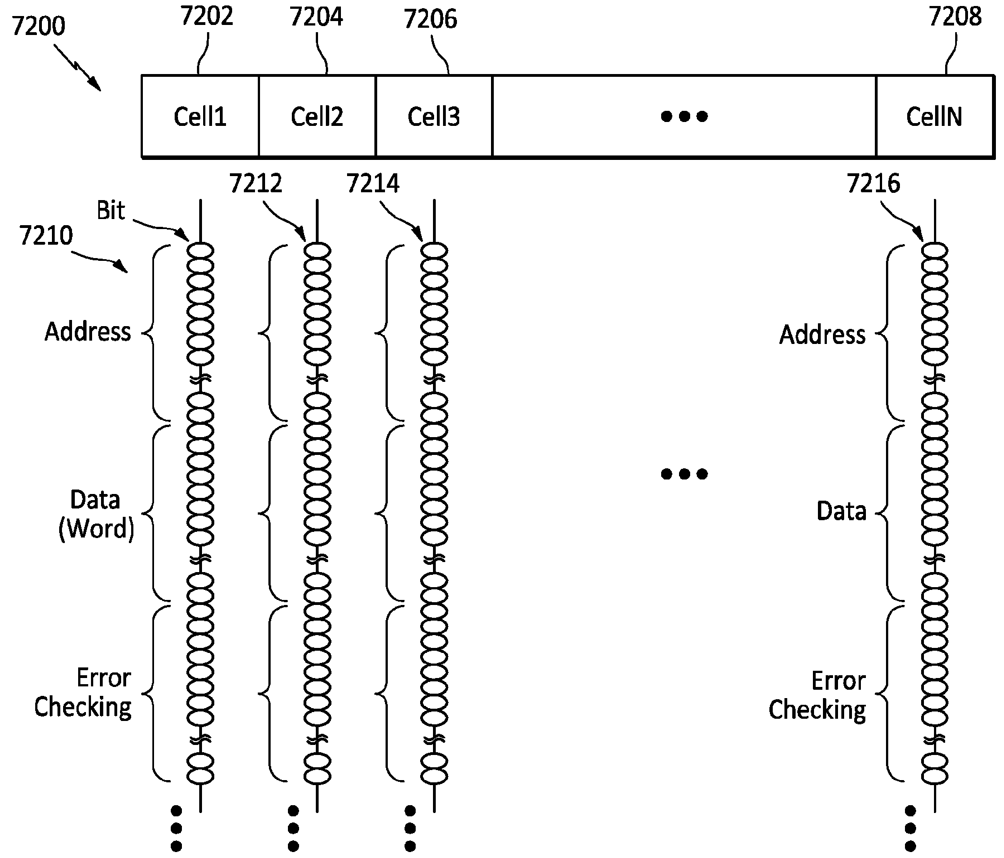}
\caption{Illustration of a collection of Iridia's NMC~\cite[Fig. 72]{Iridia19}}
\label{fig:array}
\end{figure}

Such systems can be modeled as collections of two dimensional binary arrays, which are referred to as a \textit{DNA storage arrays}, whereby each row of the array represents a particular NMC and the sequence of bits in the row represents the logical information contained within that NMC. An illustration of such a system architecture is being shown in Figure~\ref{fig:array} where each cell consists of an individual strand of DNA and this data can be partitioned into information pertaining to the address of the NMC, the data itself, and additional data such as ECC.

One of the key challenges, and one which is essential to making emerging DNA storage architectures practically viable, is constructing error correcting coding schemes. The design of such schemes not only further increases the already exceptional durability of DNA storage, but it also enables greater speeds and flexibility when it comes to reading information back from the underlying data storage medium. In this way, the problem of reading information from a DNA storage system is no longer equivalent to the challenges faced by existing biological DNA sequencers today which is to determine (exactly) the underlying sequence of nucleotides that comprises a given strand of DNA. 

Designing error correcting coding schemes for DNA data storage systems is fundamentally different than the design of coding schemes for traditional storage media~\cite{VK04}. Traditional storage typically experiences errors in the form of substitutions whereas data stored within DNA can experience insertion, deletion, and substitution errors. Insertions can be the result of the improper blocking of some bits whereas deletions can be caused by a failure of the bit addition chemistry. On top of this, cells within the DNA storage array can be lost either partially, which occurs when a defective bit prematurely terminates the chain, or the data within a cell can be corrupted completely. Initial experiments have reported error rates as high as 10\%~\cite{CKZEPK19}.
Despite these high numbers, we are unaware of any existing works that attempt to construct coding schemes for systems that can be modeled as DNA storage arrays, and this work represents the first effort towards the development of such a system.

Although the observed error rates are high, there appear to be certain commonly occurring error patterns, which may be useful in the design of future coding schemes. In particular, when a large number of errors occur, often only the tail end of the strand is affected. For shorthand, we say that an \emph{$e$-tail-erasure} (TE for short) has occurred if we are unable to recover the last $e$ symbols of the strand stored in the cell.

In the realm of error correction, we envision coding schemes that provide two levels of protection. The first level of protection corrects errors that may occur within each NMC, including insertions, deletions, and substitutions in DNA strands whereas the second level addresses more severe corruption. We propose a new class of codes, which we refer to as \emph{tail erasure codes} (\emph{TE codes} for short), for the second level of our envisioned two-level scheme. 
Our interest in codes with this structure is motivated by the fact that most of the errors that we are interested in correcting take the form of deletions and in particular as a burst of deletions. Since we know both the number of deletions that have taken place along with the location, we can treat the burst of deletions as a burst of erasures, which in general is a much easier problem to tackle than burst deletion correction.

The $e$-tail-erasure model that we study in this paper has also applications to the \emph{slow-fading channel} as suggested in~\cite{PascalAshwin2007} and the \emph{reliable-to-unreliable channel} as described in~\cite{zhou2016bch}. 
According to the reliable-to-unreliable channel model, data is transmitted over $n$ channels in a distributed way, and each channel switches from reliable transmission into an unreliable one. This switch happens during the transmission, and the position where it occurs may be different in every channel. The rest of the transmission, since the moment where it became unreliable and until the end, is considered as an erasure. Therefore, each of these channels can be modeled as a row in a two dimensional array, where the last $e$ bits of each row are considered as erasures since the channel became unreliable, and this results in a TE.

In this work, we initiate the study of TE codes. Our main result, which appears in Section~\ref{sec:TE codes}, is the development of novel error-correcting codes that we show in many cases are nearly optimal. We also consider a generalization of our problem to account for the setup where both deletions and TEs may occur and derive codes and bounds for this setup as well.  
This paper is organized as follows. In Section~\ref{sec:defs} we formally introduce our problem statement and discuss related work. In Section~\ref{sec:TE codes} we present a general construction for TE codes. Sections~\ref{sec:(t,s)-del-corr} and \ref{sec:(t,1,e)-TED-corr} extend our problem to include the scenario where in addition to TEs, deletions are allowed to occur. In Section~\ref{sec:Bounds}, bounds are derived for both TE codes and codes capable of correcting TEs and deletions. Finally, Section~\ref{sec:conclusions} concludes the paper.

\section{Notations, Definitions, Problems Statement, and Related Work}
\label{sec:defs}
Throughout this paper, let $\mathbb{F}_2$ be the field of size two, denote $[n]=\{1,\dots,n\}$ and additionally let $\mathbb{Z}_{\ge 0}=\{m\in\mathbb{Z}\colon m \ge 0\}$.
Also, for any vector $\bm{x}=(x_1,\dots,x_n)$, denote the subsequence of a vector $\bm{x}$, as $\bm{x}_{[i\colon j]}=(x_{i},\dots,x_{j})$.
Moreover, denote a binary array as $\bm{X}=(\bm{x}_1,\dots,\bm{x}_n)\in{\mathbb{F}_2}^{n\times L}$, where $\{\bm{x}_i\}_{i=1}^{n}$ are binary row vectors of length $L$, which means $\bm{x}_i=(x_{i,1},\dots,x_{i,{L}})\in\mathbb{F}_2^{L}$ for any $i \in [n]$. 
Finally, for $\bm{x}=(x_1,\dots,x_n)\in\mathbb{Z}_{\ge 0}^n$, let $\|\bm{x}\|_{1}=\sum_{i=1}^n x_i$, and $\|\bm{x}\|_{\infty}=\max\{x_1,\ldots,x_n\}$.
Next, we lay the goals of this paper and the error models we study.

\subsection{Definitions of the Error Models}
\label{sec:defs:models}
If it is not possible to recover the last bits from a row in the array, it is said that a \emph{tail-erasure} has occurred in this row. In this case, the row vector $\bm{x}_i\in\mathbb{F}_2^{L},~i \in [n]$, suffers from an \emph{$e$-row-tail-erasure} if the bits $\{x_{i,L-e+1},\dots,x_{i,L}\}$ are all erased. This definition is generalized as follows.
\begin{definition}\label{def:te}
Let $\bm{X}\in\mathbb{F}_2^{n\times L}$. It is said that $\bm{X}$ suffers from an \emph{$e$-tail-erasure}, for a positive integer $e$, if there exists a positive integer $t\le n$, positive integers $e_1,\dots,e_t$, such that $e_1+\dots+e_t=e$, and $t$ distinct row indices $i_1,i_2,\ldots,i_t$, such that for $j \in [t]$, the $i_j$-th row in $\bm{X}$ suffers an $e_j$-row-tail-erasure.
\end{definition}
For shorthand, this paper will refer to an $e$-tail-erasure simply as an \emph{$e$-TE}.
A code $\cC\subseteq\mathbb{F}_2^{n\times L}$ capable of correcting any $e$-tail-erasure is referred as an \emph{$e$-TE code}, or a \emph{TE code} when $e$ is not specified.

Subsequently, our model yields a definition to a distance function, which corresponds with the definition of the \emph{$m$-metric}, as presented first in~\cite{rosenbloom1997codes} and then further studied by~\cite{dougherty2002maximum,raviv2020hierarchical,jain2013}.
Most of these papers focus on fields of large cardinality, and when reduced to the binary field, their constructions work for a very small number of rows, $n$. 

In our paper we study specifically binary codes for arbitrary $n$, and present a new construction method, based on parity check matrices designed for this metric. This yields a new family of explicit codes, which in many cases is also better than the previous constructions in terms of the number of redundancy bits.
This will be discussed more in detail in Sections~\ref{sec:rwork}, and~\ref{sec:TE codes}.
The first goal of the paper is as follows.

\begin{problem}\label{problem1:TE}
Construct optimal TE codes.
\end{problem}

The results for this problem  are presented in Sections~\ref{sec:TE codes} and~\ref{sec:Bounds:TE Upper Bounds} and are summarized in Table~\ref{Summary-TE-Table}.
Next, it is said that an array $\bm{X}\in\mathbb{F}_2^{n\times L}$ suffers from a \emph{$(t,s)$-deletion} if at most $t\le n$ rows suffer at most $s\le L$ arbitrary deletions each. 
A code that can correct any $(t,s)$-deletion is referred to as a \emph{$(t,s)$-DC} or a \emph{DC code} for short when $t,s$ are unspecified. 

The second goal of this paper is summarized in the next problem.
\begin{problem}\label{problem2:DC}
Construct optimal DC codes.
\end{problem}

The results for this problem  are presented in Sections~\ref{sec:(t,s)-del-corr} and~\ref{sec:Bounds:DC Upper Bounds}.
Finally, this paper studies the combined model of TEs and deletions. More specifically, a \emph{$(t,s,e)$-tail-erasure-deletion}, abbreviated as a \emph{$(t,s,e)$-TED}, is a combination of an $e$-TE and a $(t,s)$-deletion. 
It can be shown that the order of occurrence between the TEs and deletions in each row does not matter, and thus for simplicity it is assumed that first the $e$-TE has occurred, which is then followed by the $(t,s)$-deletion.

However, it is worth mentioning that since the output of this channel is just an array with some truncated rows, we cannot distinguish between the scenarios of a TE and an arbitrary deletion in a given row, based on the received row's length alone. For instance, assume that in a $(1,1,1)$-TED model, one receives an array with the first two rows each missing one bit. Then, any one of the following may hold: (i) The first and second rows each experience a single TE, (ii) The first row experiences a TE and the second row does not, or (iii) The second row experiences a TE and the first does not.

A code that can correct any $(t,s,e)$-TED is a \emph{$(t,s,e)$-TED code}, or a \emph{TED code} when $t,s,e$ are unspecified. 
The third goal of the paper is described next.
\begin{problem}\label{problem3:TED}
Construct optimal TED codes.
\end{problem}

The results for this problem  are presented in Sections~\ref{sec:(t,1,e)-TED-corr} and~\ref{sec:Bounds:TED Upper Bounds}.

\subsection{Related Work}\label{sec:rwork}
Several previous works have proposed models and coding schemes for the emerging DNA storage system.
In \cite{LSWY19}, the authors modeled the DNA storage system as an ordered set of sequences of a certain length where each sequence can experience insertions, deletions, and substitutions. Other works such as \cite{IC17} and \cite{YKGM18} considered the problem of developing constrained codes that result in input sequences that are less likely to experience errors that may result from the DNA storage channel. Recently, \cite{PJSF20} developed codes for DNA storage that protect against errors but are also amenable to a broad class of coding constraints. 

The motivation for our work stems from its connection to storage systems that fall under the DNA storage array where DNA molecules are stored within nano-memory cells such as~\cite{Iridia19} rather than as an unordered multiset. Under this setting, when a particular cell is severely corrupted the errors manifest themselves as erasures that occur typically at the very end of the strand.
As discussed in Section~\ref{sec:introduction}, one can model each cell as a row of a binary array, and therefore the entire system can be modeled as an $n\times L$ binary array, where $n$ is the number of cells, and $L$ is the length of the data in the strand that is saved in each cell. Moreover, since the cells are ordered by hardware design, there is no need to reserve bits to index the strands as in other DNA systems. 

This model, where it is assumed that the indices are given, can also be applied to another common model of DNA storage systems, where all the information strands are stored in one container, which results a loss of ordering. Therefore, in this unordered information model, one usually dedicates part of the information for storing indices, as was suggested by~\cite{LSWY20} and studied by~\cite{Hengjia22,Boruchovsky23,Sima21} and more.

Since it can either be assumed that the indices are error-free, or that there is a dedicated ECC for the indices, it can be verified easily that the final resulted read of the strands in the unordered model, after ordering and removing the indices, is a binary array as in our model.

The work on TE codes is strongly related to many works on codes over either the so-called $m$-metric (also called the \emph{RT-metric} in other papers), where $m$ corresponds to the number of rows in the array, which was introduced first in~\cite{rosenbloom1997codes}.
The $m$-metric is found to be useful in the case of evaluating codes that correct erasures in arrays, which occur at the end of each column.
Therefore, it is easy to verify, using array transposition, that TE codes are equivalent to codes in the $m$-metric. Another relevant metric is the so-called \emph{$\rho$-metric}, that was first studied in~\cite{dougherty2002maximum}, where the metric corresponds to erasures that occur at the beginning of each row. 

Moreover, in Section~\ref{sec:TE codes:TE metric}, we define our \emph{$\rho_{\text{TE}}$-metric} for evaluating TE codes. However, as will be discussed in this section, the $m$-metric, the $\rho$-metric and the $\rho_{\text{TE}}$-metric are all equivalent in terms of metric spaces isomorphism. 

Lastly, we note that we prefer to define our new metric although it is equivalent to the previous one, since it better fits the error patterns studied in our paper, and will make the reading easier.

Next, a discussion is given, regarding the current state of the art results under the mentioned metrics.
In~\cite{raviv2020hierarchical,dougherty2002maximum} the case of erasures at the beginning of the row was considered, and in~\cite{rosenbloom1997codes,jain2013,zhou2016bch} it was at the end of the columns.
Moreover, in~\cite{rosenbloom1997codes,dougherty2002maximum,jain2013,raviv2020hierarchical} mostly the case of large fields was considered, and several bounds were found, as will be described in Section~\ref{sec:Bounds:TE Upper Bounds}.
Also, two results of~\cite{raviv2020hierarchical} for $q=2$ can be applied to Problem~\ref{problem1:TE}, but the problem the authors solved is more strict, in terms of linearity.
Their model treats every row as a symbol in $\mathbb{F}_{q^{L}}$ and requires linearity of the length-$n$ codewords over $\mathbb{F}_{q^{L}}$.
Therefore, under our model, our construction is less restrictive, since linearity demands are only over $\mathbb{F}_q$.
Lastly, in~\cite{zhou2016bch} a BCH code for the RT-metric is presented, and therefore a construction for a binary TE code can be derived where the length of each row is a power of two. However, their construction has concrete results only for TE codes of size $3\times 4$, and for the general case there is no explicit lower bound on the cardinality of such codes. Therefore, one cannot derive a proper comparison between their results and the results in our paper.

Another related area of research involves the design of Universally Decodable Matrices (UDM), which is originated from the slow-fading channel problem, such as in~\cite{PascalAshwin2007,raviv2020hierarchical}. Specifically, in~\cite{raviv2020hierarchical} a construction is given for the case where the rows are seen as a symbol over $\mathbb{F}_{q^L}$ and $q>n$, which does not generalize into the binary array model in this paper. Moreover, it is proved in~\cite[Theorem 4]{PascalAshwin2007} that a necessary condition for the existence of $\{A_i\}_{i=0}^{n-1}$ UDM over $\mathbb{F}_q$ is that $q>n$, and therefore any construction based strictly on UDM, cannot be used in the binary case.

We note that our problem bears resemblance to the problem of coding for segmented edit channels~\cite{LM09},~\cite{AVF17}. However, unlike the segmented edit model, we assume that the location of the tail-erasure is also known.

Nevertheless, we also face the model of $(t,s)$-deletions, that can be seen as a segmented edit channel, where at most $t$ segments are erroneous and the separation to segments is given.

Our problem is also reminiscent of constructing unequal error protection (UEP) codes \cite{MW67}, \cite{ZX05}, \cite{BK81}. UEP codes are codes with the property that some information bits are more protected against a greater number of errors than other information symbols. Under the UEP model, each coordinate, say $i$, in a codeword is pre-assigned a protection which is referred to as $f_i$. Under this setting, if $f$ errors occur in the underlying codeword, then we can determine the value of any coordinate $j$ in the original codeword if $f_j \geq f$ regardless of whether the original codeword can be determined. One of the primary motivations for the development of UEP codes was to ensure that when errors occur, the values of coordinates with higher protection levels $f_i$ can still be recovered \cite{MW67}. As is the case across many existing storage solutions \cite{lin2004error}, our interest will be in the development of codes that ensure that so long as the number of errors that occur is below a certain threshold, all errors are correctable.

As will be discussed in the next section, the distance metric of interest for constructing TE codes is fundamentally different in the sense that our schemes aim to recover the entire codeword under the setting where the locations of the errors are non-independent and satisfy certain spatial properties in the underlying arrays.

\section{Tail-Erasure Codes}
\label{sec:TE codes}
In this section, constructions of TE codes, and also linear TE codes, are presented. The main result of the section is in Section~\ref{sec:TE codes:Construction}, where a construction of a linear $e$-TE code when $e \leq L$ is presented. Moreover, several results for the case where $e>L$ can be found in Section~\ref{sec:TE-codes-large-e}.

\subsection{The TE Distance}
\label{sec:TE codes:TE metric}
The focus of this subsection is introducing a suitable distance function for TE codes, while also providing a definition of a TE pattern. 
We begin with the definition of the $\rho_{\text{TE}}$ distance, as follows.

\begin{definition}\label{def:rho-TE-distance}
    Let $\bm{X}\!=\!(\bm{x}_1,\!\dots\!,\bm{x}_n),\bm{Y}\!=\!(\bm{y}_1,\!\dots\!,\bm{y}_n)\in\mathbb{F}_q^{n\!\times\!L}$. Then, for every $j\in[n]$,
    \begin{align*}
        \rho_{\text{TE}}(\bm{x}_j,\bm{y}_j)=\begin{cases}
            L\!-\!\min\{i \colon x_{j,i} \ne y_{j,i}\}+1 &\text{ if } \bm{x}_j\ne\bm{y}_j \\
            0 &\text{ if } \bm{x}_j=\bm{y}_j
        \end{cases}~.
    \end{align*}
    And,
    \begin{align*}
        \rho_{\text{TE}}(\bm{X},\bm{Y}) = \sum_{j=1}^{n} \rho_{\text{TE}}(\bm{x}_j,\bm{y}_j)~.
    \end{align*}
\end{definition}

This distance function is closely related to the $\rho$-metric, which was introduced in~\cite{dougherty2002maximum}, while noting that the difference between the definitions is that $\rho(\bm{x}_j,\bm{y}_j)=\max\{i \colon x_{j,i} \ne y_{j,i}\}$ if $\bm{x}_j \neq \bm{y}_j$, and one can easily find the isomorphism between these two distance definitions, by reversing the order of each row.
Moreover, it can be deduced that the $\rho_{\text{TE}}$-distance is a metric and from now on we will refer to it as the $\rho_{\text{TE}}$-metric or the TE-metric.
The relation to the RT-metric is by transposing the array, and therefore, in terms of code parameters, all of the above mentioned metrics give equivalent codes and a comparison between the codes that are derived in each metric will be given at the end of Section~\ref{sec:TE codes:Construction}.

Next, a mathematical definition of a \emph{TE pattern} is given.
This definition resembles with the ones in~\cite{PascalAshwin2007,raviv2020hierarchical}, but is more suitable for the notations of this paper.
\begin{definition}\label{def:TE-patterns-set}
For $e,L,n$, the \emph{TE-pattern set} is defined to be 
$$P(e,L,n) = \left\{\bm{p}\in\mathbb{Z}_{\ge 0}^{n} \colon \|\bm{p}\|_{1}\le e, \|\bm{p}\|_{\infty}\le L\right\}~.$$
\end{definition}
In other words, $P(e,L,n)$ is the set of all length-$n$ vectors $\bm{p}$, which are called TE patterns, where the sum of entries in $\bm{p}$ is at most $e$ and the value of each entry in $\bm{p}$ is at most $L$. A pattern $\bm{p}\in P(e,L,n)$ will be used next as an indicator, such that the $i$-th entry in $\bm{p}$ represents the number of erased bits in the $i$-th row of the array.
Following this, the concept of a $\bm{p}$-pattern-TE is introduced.
\begin{definition}\label{def:p-pattern-TE}
Let $\bm{X}\in\mathbb{F}_2^{n\times L}$ and $\bm{p}\in P(e,L,n)$. Then, the \emph{$\bm{p}$-pattern-TE} of $\bm{X}$, denoted by $\bm{X}^{(\bm{p})}\in(\mathbb{F}_2\cup\{?\})^{n\times L}$, is defined as follows: 
\begin{align*}
x^{(\bm{p})}_{i,j}=\begin{cases}
? &\text{if } p_i > 0 \text{ and } j \geq L-p_i + 1\\
x_{i,j} &\text{else.}
\end{cases}~.
\end{align*}
\end{definition}
Conceptually, $\bm{X}^{(\bm{p})}$ represents the array that is obtained by replacing the last $p_i$ symbols in the $i$-th row of $\bm{X}$ with the erasure symbol "$?$". An example is given next.
\begin{example}
Let 
$\bm{X}=\begin{bmatrix}
1 & 0 & 1\\
0 & 0 & 1
\end{bmatrix}~,\quad\bm{p}=(2,1)$.
Then, 
$\bm{X}^{(\bm{p})}=\begin{bmatrix}
1 & ? & ?\\
0 & 0 & ?
\end{bmatrix}$,
is a $\bm{p}$-pattern-TE of $\bm{X}$.
\end{example}
Next, we provide a claim that connects Definitions~\ref{def:TE-patterns-set} and~\ref{def:p-pattern-TE} to the $\rho_{\text{TE}}$-metric definition.
\begin{claim}\label{claim:rho-metric-is-TE-metric}
Let $\bm{X},\bm{Y}\in\mathbb{F}_2^{n\times L}$. Then,
\begin{equation}
    \rho_{\text{TE}}(\bm{X},\bm{Y}) = \min_{\bm{p}\in P(nL,L,n)} \left\{ \|\bm{p}\|_{1} \colon \bm{X}^{(\bm{p})} = \bm{Y}^{(\bm{p})} \right\}. \nonumber
\end{equation}
\end{claim}
\begin{IEEEproof}
    By definition, for every row $j\in[n]$, the distance $\rho_{\text{TE}}(\bm{x}_j,\bm{y}_j)$ is exactly the minimal number of erasures required in the end of both $\bm{x}_j$ and $\bm{y}_j$ so they appear the same. In other words, this is the value of $p_j$ for some vector $\bm{p}=(p_1,\dots,p_n)$, such that $(\bm{X}^{(\bm{p})})_j = (\bm{Y}^{(\bm{p})})_j$, where $(\bm{X}^{(\bm{p})})_j$ denotes the $j$-th row of $\bm{X}^{(\bm{p})}$ and $(\bm{Y}^{(\bm{p})})_j$ denotes the $j$-th row of $\bm{Y}^{(\bm{p})}$.
    Thus, a vector $\bm{p}$ that obtains the minimum $\|\bm{p}\|_{1}$, for which $\bm{X}^{(\bm{p})} = \bm{Y}^{(\bm{p})}$ as in the left hand side, is constructed in a way such that every entry $p_j$, represents also the minimal number of erasures required in the end of both $\bm{x}_j$ and $\bm{y}_j$, such that $(\bm{X}^{(\bm{p})})_j = (\bm{Y}^{(\bm{p})})_j$, or otherwise it results a contradiction to the minimality, since one can choose a smaller value for the $j$-th entry of $\bm{p}$. 
    Hence, for the above minimal $\bm{p}$ it holds that
    $$\|\bm{p}\|_{1}=\sum_{i=1}^{n}p_i = \sum_{j=1}^{n} \rho_{\text{TE}}(\bm{x}_j,\bm{y}_j) = \rho_{\text{TE}}(\bm{X},\bm{Y})~,$$
    which completes the proof.
    
\end{IEEEproof}
A short example of computing the $\rho_{\text{TE}}$-metric by using a TE pattern is given next.
\begin{example}\label{ex:eho-TE-dist}
Let $\bm{X},\bm{Y}\in\mathbb{F}_2^{2\times 3}$ be defined as follows,
$$\bm{X}=\begin{bmatrix}
1 & 0 & 1\\
0 & 0 & 1
\end{bmatrix}~,\quad
\bm{Y}=\begin{bmatrix}
1 & 0 & 0\\
0 & 1 & 1
\end{bmatrix}~.$$
It can be readily verified that $\rho_{\text{TE}}(\bm{X},\bm{Y})=3$ using $\bm{p}=(1,2)$, or using Definition~\ref{def:rho-TE-distance}.
\end{example}

Compared to the Hamming distance, note that under this model, erasures do not independently contribute to the $\rho_{\text{TE}}$-distance between two codewords. 
For example, consider the case where $\bm{X}$ is as in Example~\ref{ex:eho-TE-dist}. 
Then, the vectors 
$$\bfY = \begin{bmatrix}
1 & \bm{1} & \bm{0}\\
0 & 0 & \bm{0}
\end{bmatrix},\quad\bfZ = \begin{bmatrix}
1 & \bm{1} & 1\\
0 & 0 & \bm{0}
\end{bmatrix}~,$$
are each at the same TE distance (of three) to $\bfX$ despite the fact that the set of coordinates in which $\bfX$ and $\bfZ$ differ is a subset of the set of coordinates where $\bfX$ and $\bfY$ do not agree.

Finally, a definition of a minimum distance of a code under the $\rho_{\text{TE}}$ distance follows.
\begin{definition}\label{def:TE code min dist}
Let $\cC\subseteq\mathbb{F}_2^{n\times L}$ be a TE code. Then, the \emph{minimum $\rho_{\text{TE}}$-distance of $\cC$} is defined as follows
\begin{equation}
    \rho_{\text{TE}}(\cC) = \min_{\bm{X},\bm{Y}\in\cC\colon \bm{X}\ne\bm{Y}} \rho_{\text{TE}}(\bm{X},\bm{Y})~. \nonumber
\end{equation}
\end{definition}
Next, denote by $(n \times L,M,e+1)_{\text{TE}}$ a code of cardinality $M$, and minimum $\rho_{\text{TE}}$-distance of $e+1$ over binary arrays of length $n\times L$. Then, a basic theorem regarding the minimum $\rho_{\text{TE}}$-distance of a code is given. This theorem illustrates one of the similarities between the properties of classical codes under the Hamming metric, and the TE codes in our paper, as we show that an $(n \times L,M,e+1)_{\text{TE}}$ code can correct $e$ erasures.
Although this theorem is probably known, we present it here for the sake of completeness.

\begin{restatable}{theorem}{tecodeiff}\label{thm:te-code-iff}
A code $\cC\subseteq \mathbb{F}_2^{n\times L}$ is an $e$-TE-correcting code if and only if $\rho_{\text{TE}}(\cC)\ge e+1$.
\end{restatable}

The proof of Theorem~\ref{thm:te-code-iff} is provided in the appendix.
Furthermore, Theorem~\ref{thm:te-code-iff} indicates that the constructions presented later in the paper are applicable to any type of TE pattern.

\subsection{Tail-Erasure Linear Codes}
\label{sec:TE codes:linear codes}
Next, linear TE codes are constructed.
First, the notation of a parity check matrix for linear codes over arrays under the $\rho_{\text{TE}}$-distance is presented.
\begin{definition}
    Let $\cC\subseteq\mathbb{F}_2^{n\times L}$ be a linear code. A \emph{TE parity check matrix} of $\cC$ is a three-dimensional array  $\bm{\cH}\in\mathbb{F}_2^{r\times n\times L}$,
    
    \begin{equation*}
        \bm{\cH} = 
        \begin{bmatrix}
            \bm{h}_{1,1} & \dots & \bm{h}_{1,L} \\
            \vdots & \ddots & \vdots \\
            \bm{h}_{n,1} & \dots & \bm{h}_{n,L}
        \end{bmatrix}~, \bm{h}_{i,j}\in\mathbb{F}_2^{r}~,
    \end{equation*}
    
    where it holds that
    
    \begin{equation*}
        \bm{X}=\left( x_{i,j} \right)_{i \in [n], j \in [L]}\in\cC \iff \sum_{i=1}^{n}\sum_{j=1}^{L}x_{i,j}\bm{h}_{i,j}=\bm{0}~.
    \end{equation*}
    
\end{definition}

Denote $\bm{\cH}^{*}$ as the set of vectors, which are the entries of $\bm{\cH}$, i.e., $\{\bm{h}_{1,1}, \ldots, \bm{h}_{n,L}\}$. Then, $\dim(\cC) = n L - \text{rank}(\bm{\cH}^{\textcolor{blue}{*}})$. Moreover, refer to a linear TE code $\cC\subseteq\mathbb{F}_2^{n\times L}$, as an $[n\times L,k,d]_{\text{TE}}$ code, where $k=\dim(\cC)$ and $d=\rho_{\text{TE}}(\cC)$.

Next, an expression of $\rho_{\text{TE}}(\cC)$ for linear codes will be developed, using several preliminary definitions. 
First, let 
$$\cJ(\bm{\cH},\bm{p}) = \left\{\left\{ \bm{h}_{i,j} \in \mathbb{F}_2^{r} \colon p_i > 0, \text{ and } j \geq L - p_i + 1   \right \}\right \}~,$$ 
where $\bm{p}\in P(e,L,n)$.
This multiset of columns represents the columns of $\bm{\cH}$ which are located in the matching entries of the erasure pattern $\bm{p}$. 
The reason for choosing the multiset over a set, is the fact that there could be a repetition of columns in $\bm{\cH}$, and it clearly affects the linear dependency of the multiset.
\begin{example} \label{ex:I(C,p)}
    Let $H=[\bm{h}_1,\dots,\bm{h}_7]$ be a parity check matrix of the $[7,4,3]$ Hamming code. Let
    
    \begin{align*}
        \bm{\mathcal{H}}=&
        \begin{bmatrix}
        \bm{h}_{2} & \bm{h}_{1} \\
        \bm{h}_{3} & \bm{h}_{2} \\
        \bm{h}_{4} & \bm{h}_{3} \\
        \bm{h}_{5} & \bm{h}_{4} \\
        \bm{h}_{6} & \bm{h}_{5} \\
        \bm{h}_{7} & \bm{h}_{6} \\
        \bm{h}_{1} & \bm{h}_{7}
        \end{bmatrix}\in \mathbb{F}_2^{3\times 7\times 2},~
        \bm{p}=\begin{pmatrix}1\\0\\0\\0\\0\\0\\2\end{pmatrix}.
    \end{align*}
    
    Then, $\cJ(\bm{\cH},\bm{p})=\{\{\bm{h}_{1}, \bm{h}_{1},\bm{h}_{7}\}\}$. 
    Notice that there are repetitions of columns and therefore this multiset is linearly dependent. This is one of the cases where a TE pattern cannot be corrected, and this observation will be generalized in Claim~\ref{TE-min-dist-claim}. 
\end{example}
In addition to the above, define the \emph{TE-weight} of $\bm{X}$ as $w_{\text{TE}}(\bm{X})\!=\!\rho_{\text{TE}}(\bm{X},\bm{0})$.
It is possible to show that
$\rho_{\text{TE}}(\bm{X},\bm{Y})\!=\!\rho_{\text{TE}}(\bm{X}\!-\bm{Z}\!,\bm{Y}\!-\bm{Z})$ for any $\bm{Z}$.
Then, if $\bm{Z}\!=\!\bm{Y}$, $\rho_{\text{TE}}(\bm{X}\!,\bm{Y})\!=\!w_{\text{TE}}(\bm{X}\!-\!\bm{Y})$.

In~\cite{zhou2016bch} the authors indicated another similarity between the Hamming distance and the $\rho_{\text{TE}}$-distance, which is,
\begin{equation}
    \rho_{\text{TE}}(\cC) = \min_{ \bm{X} \in \cC \setminus \{\bm{0}\} } w_{\text{TE}} (\bm{X})~.\nonumber
\end{equation}

Lastly, a connection between the TE parity check matrix and the minimum $\rho_{\text{TE}}$-distance of its code is presented.
\begin{claim}\label{TE-min-dist-claim}
Let $\bm{\cH}$ be a TE parity check matrix of a code $\cC\subseteq\mathbb{F}_2^{n\times L}$. Then, $\rho_{\text{TE}}(\cC)$ is the largest integer $d$, such that
for any $\bm{p}\in P(d-1,L,n)$, 
the vectors in $\cJ(\bm{\cH},\bm{p})$ are linearly independent.
\end{claim}
\begin{IEEEproof}
Let $\bm{X}\in\cC$ with TE weight of $t>0$. This implies the existence of $\bm{p}\in P(t,e,n), \|\bm{p}\|_{1}=t$ such that $\cJ(\bm{\cH},\bm{p})$ is of cardinality $t$, and denote the column indices in $\cJ(\bm{\cH},\bm{p})$ as $J$. 
From the definition of TE weight, any $(i,j)\notin J$ implies $x_{i,j}=0$ and since $\bm{X}$ is a codeword,
\begin{align*}
    \sum_{i=1}^{n}\sum_{j=1}^{e}\bm{h}_{i,j}x_{i,j} = \sum_{(i,j)\in J}\bm{h}_{i,j}x_{i,j} =\bm{0} \nonumber~,
\end{align*} 
which implies $\cJ(\bm{\cH},\bm{p})$ is a linearly dependent multiset by definition.
Conversely, let $\cJ^{\prime}(\bm{\cH},\bm{p}^{\prime})$ be a multiset of $t^{\prime}$ linearly dependent columns of $\bm{\cH}$, for some $\bm{p}^{\prime}\in P(t^{\prime},e,n)$, and denote the column indices in $\cJ^{\prime}(\bm{\cH},\bm{p}^{\prime})$ as $J^{\prime}$, i.e., 
$\sum_{(i,j)\in J^{\prime}}\alpha_{i,j}\bm{h}_{i,j} =\bm{0}$,
which implies the existence of a codeword $\bm{Y}$ which is zero in every entry except the indices in $J^{\prime}$, thus $0<w_{\text{TE}} (\bm{Y})\le t^{\prime}$, and then $d\le t^{\prime}$, so we can conclude that there are no patterns $\bm{p}^{\prime}\in P(t^{\prime},L,n)$, for $t^{\prime}<d$, such that the vectors in $\cJ(\bm{\cH},\bm{p}^{\prime})$ are linearly dependent.

\end{IEEEproof}

\subsection{Linear Tail Erasure Code Construction}
\label{sec:TE codes:Construction}
In this subsection, we give a construction for linear codes where $e\le L$. Note that if $e \leq L$, then the first $L-e$ bits of a row cannot be erroneous. 
Thus, the subarray of size $n \times (L-e)$ of any word of size $n \times L$, taken as its first $L-e$ columns is error-free. Therefore, one can use the constructions in this subsection for the last $e$ columns, and then prepend any first $L-e$ columns subarray to it, without affecting the bits of redundancy or the TE-correcting ability of the code.
Hence, unless stated otherwise, it is assumed throughout this subsection $L=e$ so that the size of our arrays are $n \times \left( L=e\right)$.
In Section~\ref{sec:TE-codes-large-e} a discussion on codes where $e > L$ is given, and also a construction of a linear code for this case will be given in Section~\ref{sec:TE-codes-large-e}.

First, since one can take a parity check matrix and construct a generator matrix from it, we can derive an efficient (polynomial-time) encoding scheme. 
For decoding, since the locations of the errors are known under this model, a system of linear equations can be used to correct erasures, using the same parity check matrix. Therefore, an efficient decoding scheme is also implied.
Following the above discussion, a construction of an $[n\times(d-1),k,d]_{\text{TE}}$ code, $\cC$, is presented, using a TE parity check matrix.
\begin{construction}\label{construction:linear-TE-code}
Let $d$ be an odd positive integer and denote $t\!=\!\frac{d-1}{2}$. Also, let $\cC_B$ be a binary $[nt,k_B,d]$ error-correcting code, referred as the \emph{base code}, with a parity check matrix $H_B =[\bm{h}_{1}\,\bm{h}_{2} \cdots \bm{h}_{nt}]$. Then, 
\begin{align*}
    \bm{\mathcal{H}}=
    \left[\begin{array}{ccc|ccc}
        \bm{h}_{1} & \cdots & \bm{h}_{t} & \bm{h}_{2t} & \cdots & \bm{h}_{t+1} \\
        \bm{h}_{t+1} & \cdots & \bm{h}_{2t} & \bm{h}_{3t} & \cdots & \bm{h}_{2t+1} \\
        \vdots & \ddots & \vdots & \vdots & \ddots & \vdots \\
        \bm{h}_{(n-1)t+1} & \cdots & \bm{h}_{nt} & \bm{h}_{t} & \cdots & \bm{h}_{1}
    \end{array}\right]~,
\end{align*}
is a TE parity check matrix of a TE-correcting code $\cC_{\text{TE}}$.
\end{construction}
\begin{remark}\label{remark:te-for-n-eq-2}
For $n=2$, this construction is degenerate, since it requires a $[2t,k_B,2t+1]$ base code.
\end{remark}

For example, the code with the TE parity check matrix $\bm{\mathcal{H}}$ in Example \ref{ex:I(C,p)} is a $[7\times 2,11,3]_{\text{TE}}$ code.
The next claim emphasizes an important property of Construction~\ref{construction:linear-TE-code}.
\begin{claim}\label{clm:aux-for-const-TE-thm}
    Let $\bm{p}\in P(e,2t,n)$, where $e\le 2t$, and $\bm{\cH}$ as in Construction \ref{construction:linear-TE-code}. Then, the multiset $\cJ(\bm{\cH},\bm{p})$ does not contain any column vector with multiplicity greater than $1$.
\end{claim}
\begin{IEEEproof}
    Assume, for the sake of contradiction, that there exists a column vector with multiplicity $2$ in $\cJ(\bm{\cH},\bm{p})$ (according to the construction it cannot have a greater multiplicity). 
    By definition, for an arbitrary $k\in[nt]$, $\bm{h}_{k}$ appears in $\bm{\cH}$ at locations $(i,j)$ and $(i-1,2t-j+1)$ (up to modulo $n$ and $2t$, respectively).
    But this implies that $\bfh_{i,j}$ and $\bfh_{i-1,2t-j+1}$ are both in $\cJ(\bm{\cH},\bm{p})$, which is not possible since $j+2t-j+1 = 2t+1>e=\|\bm{p}\|_{1}$, and $\bm{p}\in P(e,2t,n)$.
    
\end{IEEEproof}

The following presents a proof of the correctness of Construction~\ref{construction:linear-TE-code}, utilizing Claim~\ref{clm:aux-for-const-TE-thm}.

\begin{theorem}\label{thm:const_linear_TE_works}
    Let $\cC_B$ be the $[nt,k_B,d]$ base code, where $t=\frac{d-1}{2}$. Let $\cC$ be the resulting TE-correcting code from Construction \ref{construction:linear-TE-code}. Then, $\cC$ is an $[n \times 2t ,k_B+nt,d]_{\text{TE}}$ code. Furthermore, the redundancy of the code $\cC$ is $nt-k_B$.
\end{theorem}
\begin{IEEEproof}
    One can verify that this construction yields a code of length $n \times 2t$ and that $\dim(\cC)=2nt-\text{rank}(\bm{\mathcal{H}}) = 2nt-(nt-k_B) = k_B+nt$, since $\bm{\mathcal{H}}$ is constructed from the columns of $H_B$, repeated two times, and therefore $\text{rank}(\bm{\mathcal{H}}) = \text{rank}(H_B)=nt-k_B$.
    It is left to prove that $\rho_{\text{TE}}(\cC)=d$.
    Based on Claim \ref{TE-min-dist-claim}, to show that $\rho_{\text{TE}}(\cC)\ge d$, it is required to show that for any arbitrary $\bm{p}\in P(2t,2t,n)$, the  multiset $\cJ(\bm{\cH},\bm{p})$, is linearly independent.
    Since every $2t=d-1$ columns in the base code are linearly independent, the fact that $\rho_{\text{TE}}(\cC)\ge d$, immediately follows from Claim~\ref{clm:aux-for-const-TE-thm}.
    Lastly, the fact that $\rho_{\text{TE}}(\cC)=d$ can be easily verified, e.g. using the vector $\bm{p}=(2t,0,\dots,0,1)$, which results in $\cJ(\bm{\cH},\bm{p})$, which is a linearly dependent multiset.
    
\end{IEEEproof}

Construction~\ref{construction:linear-TE-code} results in codes with odd minimum $\rho_{\text{TE}}$-distance. The case of even minimum $\rho_{\text{TE}}$-distance is considered next. First, note that for minimum $\rho_{\text{TE}}$-distance 2, it is straightforward to see that a single parity bit suffices. For even minimum $\rho_{\text{TE}}$-distance greater than $2$, denote again $t=\frac{d-1}{2}$, where $d$ is odd. Then, let $\cC_B$ be the $[nt+1,k_b,d+1]$ base code, i.e., a code with even minimum (Hamming) distance $d+1$ (w.l.o.g, based on the same base code from Construction \ref{construction:linear-TE-code}, with odd minimum distance, after adding one parity bit, so that $d+1$ is even), where
$H_{B}^{*}=[\bm{h}_{1}\cdots\bm{h}_{nt+1}]$ is a parity check matrix of $\cC_B$.
Then,
\begin{align*}
    \bm{\mathcal{H}^{*}}=
    \left[\begin{array}{ccc|c|ccc}
        h_{1} & \cdots & h_{t} & h_{nt+1} & h_{2t} & \cdots & h_{t+1} \\
        h_{t+1} & \cdots & h_{2t} & h_{nt+1} & h_{3t} & \cdots & h_{2t+1} \\
        \vdots & & \vdots & \vdots & \vdots & & \vdots \\
        h_{(\!n\!-\!1\!)t\!+\!1} & \cdots & h_{nt} & h_{nt+1} & h_{t} & \cdots & h_{1}
    \end{array}\right].
\end{align*}

Similarly to Construction \ref{construction:linear-TE-code}, it can be shown that $\bm{\mathcal{H}^{*}}$ is a parity check matrix for a $[n\times (2t+1),k_b+n(t+1)-1,d+1]_{\text{TE}}$ code.
Therefore, the redundancy of such a code is $nt-k_b+1$.
In order to analyze the redundancy result of Construction~\ref{construction:linear-TE-code}, and its extension for the even minimum $\rho_{\text{TE}}$-distance case, let $R(n,d)$ be the minimum redundancy of any length-$n$ linear code with minimum (Hamming) distance $d$. 
It is known that~\cite[p. 160-161]{RothBook2006},
\begin{align*}
    R(n,d) \leq \begin{cases}
        \frac{d-1}{2}\lceil\log(n+1)\rceil, & d \text{ is odd} \\
        1+\left(\frac{d}{2}-1\right)\lceil\log(n+1)\rceil, & d \text{ is even.}
    \end{cases}
\end{align*}
Next, for an $[n\times(d-1),k,d]_{\text{TE}}$ code that can be achieved by Construction~\ref{construction:linear-TE-code}, let $R_{\text{TE}}(n,d)$ be the minimum number of redundancy bits of the code.
The next corollary considers the redundancy of Construction~\ref{construction:linear-TE-code} and provides an upper bound on the redundancy of TE-codes.

\begin{corollary}\label{cor:TE-redundancy-upper-bound}
It holds that 
\begin{enumerate}[leftmargin=*]
\item $R_{\text{TE}}(n,d) \leq R\left(\frac{n(d-1)}{2},d\right)$ and thus, when $d$ is odd,
$$R_{\text{TE}}(n,d)\leq\frac{d-1}{2}\left\lceil\log\left(n\left(d-1\right)+2\right)\right\rceil-\frac{d-1}{2}~.$$
\item $R_{\text{TE}}(n,d) \leq R\left(\frac{n(d-2) + 2}{2}, d \right)$ and thus, when $d$ is even and greater than 2,
$$R_{\text{TE}}(n,d) \leq \left(\frac{d}{2} - 1 \right)\lceil\log(n\left(d-2\right)+4)\rceil - \frac{d-4}{2}~.$$
\end{enumerate}
\end{corollary}

Note that a trivial construction of TE codes can be achieved by considering the array of size $n\times (d-1)$ as a vector of length $n(d-1)$, and then correcting $d-1$ TEs using a linear $[n(d-1),k,d]$ code.
This was presented in~\cite{zhou2016bch} as the \emph{vector construction}.
The minimal number of redundancy bits of the vector construction will be denoted by $R\left(n(d-1),d\right)$. Then, from the previous discussion along with Corollary~\ref{cor:TE-redundancy-upper-bound},
$$R\left(n(d-1),d\right) \geq R\left(\frac{n(d-1)}{2},d\right) \geq R_{\text{TE}}(n,d)~,$$
for any $n$ and odd $d$, and similarly for the even $d$ case.
More explicitly, the savings in Construction~\ref{construction:linear-TE-code} is roughly $d/2$ bits of redundancy.
Next, $[n \times (d-1) ,k,d]_{\text{TE}}$ codes are presented, which imply some additional constructive upper bounds on the redundancy of TE codes.
\begin{claim}\label{clm:TE-3-4}
    Let $n$ be a positive integer and $r=\lceil\log_2(n+1)\rceil$. Then,
    \begin{enumerate}
        \item An $[n\times 2, 2n-r, 3]_{\text{TE}}$ code exists, and thus $R_{\text{TE}}(n,3) \le \lceil\log_2(n+1)\rceil$. 
        \item An $[n\times 3, 3n-r-1, 4]_{\text{TE}}$ code exists, and thus $R_{\text{TE}}(n,4) \le \lceil\log_2(n+1)\rceil+1$.
    \end{enumerate}
\end{claim}
\begin{IEEEproof}
    Let $\cC_B$ be an $[n,n-r,3]$ base code, which is a shortening of a $(2^{r}-1)$-length Hamming code (where if $r=\log_2(n+1)$ exactly, no shortening is required), and $\cC_B^{*}$ be a $[n+1,n-r,4]$ code, which is the extended version of $\cC_B$, which one gets by adding a parity bit. Therefore, 
    \begin{enumerate}
        \item Using Theorem~\ref{thm:const_linear_TE_works}, there exists a $[n\times 2, 2n-r, 3]_{\text{TE}}$ code.
        \item From the discussion on TE codes with even $\rho_{\text{TE}}(\cC)$, there exists a $[n\times 3, 3n-r-1, 4]_{\text{TE}}$ code.
    \end{enumerate}
\end{IEEEproof}
By Claim~\ref{clm:TE-3-4} one derives an explicit upper bound on the redundancy of codes which correct 2 or 3 TEs, based on Construction~\ref{construction:linear-TE-code}.
Next, another upper bound on the redundancy of codes that correct 4 TEs is derived from linear cyclic codes, using a specific construction which is different than Construction~\ref{construction:linear-TE-code}.
\begin{claim}\label{clm:TE-5}
    Let $n>0$ be an integer, $m=\lceil\log_2(n+5)\rceil$ and $r=2m+1$. Then, there exist a $[n\times 4, 4n-r, 5]_{\text{TE}}$ code.
\end{claim}
\begin{IEEEproof}
    For this proof, a construction is given, similar to Construction~\ref{construction:linear-TE-code}, but with different ordering of the entries.
    Let $\cC_B$ be an $[n+4,n+4-r,\ge 6]$ base code, which is a shortening (if needed) of the $[2^m-1, 2^m-2m-2,\ge 6]$ cyclic code in~\cite[Example 8.10]{RothBook2006}, with a parity check matrix $H=[\bm{h}_{1}~\bm{h}_{2} \cdots \bm{h}_{n+4}]$.
    Then,
    \begin{align*}
    \bm{\cH}=
    \left[\begin{array}{cc|cc}
        \bm{h}_{n+4} & \bm{h}_{n+3} & \bm{h}_{2} + \bm{h}_{3} & \bm{h}_{1} \\
        \bm{h}_{n+4} & \bm{h}_{n+3} & \bm{h}_{3} + \bm{h}_{4} & \bm{h}_{2} \\
        \vdots & \vdots & \vdots & \vdots \\
        \bm{h}_{n+4} & \bm{h}_{n+3} & \bm{h}_{n+1} + \bm{h}_{n+2} & \bm{h}_{n} \\
    \end{array}\right]~,
    \end{align*}
    is a TE parity check matrix of $\cC$, an $[n\times 4, 4n-r, 5]$ TE-correcting code.
    It is clear why the length is $n\times 4$ and the dimension is $4n-r$, since the redundancy stays the same as in the base code.
    What is left to prove is that $\rho_{\text{TE}}(\cC)=5$, and the method will be to show that for any $\bm{p}\in P(4,4,n)$, the vectors in $\cJ(\bm{\cH},\bm{p})$ are linearly independent.
    It can be verified that every vector $\bm{p}$ that does not contain exactly two entries of value $2$ satisfies the condition of $\cJ(\bm{\cH},\bm{p})$ being linearly independent, since it will be a sum of at most $5$ distinct vectors of $H$, which is a parity check matrix of a code with minimum Hamming distance of at least $6$.
    
    Therefore, observe next on the case where there are two entries of $\bm{p}$, say $\bm{p}_i,\bm{p}_j$, where $i<j$, such that $\bm{p}_i=\bm{p}_j=2$. 
    If $j-i \le 2$, the sum of elements in $\cJ(\bm{\cH},\bm{p})$ is again only at most $5$ distinct vectors of $H$.
    So the last case one needs to check is where $j-i>2$, therefore $\cJ(\bm{\cH},\bm{p})$ can be written as $\left\{\left\{ h_{i}, h_{i+1}+h_{i+2}, h_{j},h_{j+1}+h_{j+2}\right\}\right\}$.
    To verify those vectors are independent, observe the equation, $\alpha_1 h_{i} + \alpha_2 h_{i+1} + \alpha_3 h_{i+2} + \alpha_4 h_{j} + \alpha_5 h_{j+1} + \alpha_6 h_{j+2} = 0$, where $\alpha_1,\ldots,\alpha_6 \in \mathbb{F}_2$, and all the six elements are distinct.
    
    It is assumed that either $\alpha_1=\alpha_2=\cdots=\alpha_6=0$ or $\alpha_1=\alpha_2=\cdots=\alpha_6=1$, since there cannot be a linear dependent set of five distinct vectors.
    If those are all zero, the proof is done.
    So, assume, for sake of contradiction, that $h_{i} + h_{i+1} + h_{i+2} + h_{j} + h_{j+1} + h_{j+2} = 0$, which means there is a codeword in $\cC_B$ of Hamming weight $6$, with ones only in the entries $\{i,i+1,i+2,j,j+1,j+2\}$, denoted as $\bm{X}_1$.
    Since the code is cyclic, there exists a codeword with ones only in the entries $\{i+1,i+2,i+3,j+1,j+2,j+3\}$, denoted as $\bm{X}_2$. But then, also the sum $\bm{X}=\bm{X}_1+\bm{X}_2$ is another codeword, but the Hamming weight of $\bm{X}$ is at most $4$, which is a contradiction.
    
\end{IEEEproof}

One can verify that $2\lceil\log_2(n+5)\rceil+1<2\lceil\log_2(2n+1)\rceil$ for $n\ge 4$, if $n\notin\bigcup_{i=3}^{\infty}\{2^{i}-4,2^{i}-3,2^{i}-2,2^{i}-1\}$. This implies that for most (but not all) values of $n$, the construction in Claim~\ref{clm:TE-5} gives a tighter upper bound for $d=5$ than Construction~\ref{construction:linear-TE-code} by one bit.

To conclude, the results are summarised in Table \ref{Summary-TE-Table}, where the upper bound is the constructive one from the above discussion and the lower bounds from Section \ref{sec:Bounds:TE Upper Bounds}.
Moreover, a comparison of the results presented in this section and some previous results under the $m$-metric is given next.

In~\cite{rosenbloom1997codes}, a family of MDS codes is proved to exist, since those achieve the Singleton bound in Claim~\ref{clm:Singleton-bound}. 
Similarly to the classical MDS codes for length-$n$ vectors, for the family of array codes to fit large $n$ (and arbitrary $L$), where $n$ is the number of rows in the array, it is required to use a large field size, i.e., $q\ge n$, where $q$ is the size of the field.
Nevertheless, restricting to the binary case, one can construct a code for $n=2$, which is a $[2\times L,2L-d+1,d]_{\text{TE}}$ code, where $d$ is arbitrary, and which uses only $d-1$ bits of redundancy. 
The construction in~\cite{rosenbloom1997codes} also solves the case where $e>L$, still only for $n=2$ in the binary case, as will be discussed in Section~\ref{sec:TE-codes-large-e}.
Contrary to that, the first construction in~\cite{raviv2020hierarchical}, while also be an MDS code for $n=2$, do come with a restriction of $e\le L$. 
However, the authors give a more explicit construction of an MDS code for the $2\times L$ binary array case.
As mentioned in Remark~\ref{remark:te-for-n-eq-2}, Construction~\ref{construction:linear-TE-code} is degenerated in the $n=2$ case and therefore one cannot compare between the two constructions.

Another important family of codes, introduced in~\cite{zhou2016bch}, is the construction of BCH codes for the $m$-metric. In this paper, a construction of Reed-Solomon codes for the $m$-metric is given, using Galois-Fourier transform and Hasse derivatives, and then a BCH code is derived using alternant code~\cite[Section 5.5]{RothBook2006}.
However, there are no specific bounds mentioned, and the authors only provide a specific construction example for the case where $n=3$ and $L=4$. Therefore, a comparison between their construction and this paper is parameter-dependent. 
As for their construction for $3\times 4$ binary arrays~\cite[Table I]{zhou2016bch}, one can see that for $d\le5$ ($e\le L$), a comparison can be done using Claims~\ref{clm:TE-3-4} and~\ref{clm:TE-5}. 
In our construction, we use at most $1,2,3,6$ bits of redundancy for $d=2,3,4,5$, respectively, while their optimal choice of parameters yields at most $1,3,4,6$ bits of redundancy for $d=2,3,4,5$, respectively, therefore our construction improves some of their result, or achieves the same result for the other cases. Other types of basic general construction ideas in~\cite{zhou2016bch} are the vector construction, that we mentioned before, and the \emph{Cartesian product construction}, which will be both discussed more in Section~\ref{sec:TE-codes-large-e}, but in their paper the authors claim that these two constructions are inferior compared to their BCH codes.
In~\cite{raviv2020hierarchical}, the authors gave mostly constructions for larger fields, but two constructions are relevant for the binary case. The first was mentioned before, as an MDS code for $n=2$, and the second will be discussed next in Section~\ref{sec:TE-codes-large-e}.

\begin{table*}[t]
\centering
\caption{Bounds on the Redundancy of TE-Correcting Codes.}
\begin{tabular}{||c c c c c||}
\hline
Array Length & Minimum $\rho_{\text{TE}}$-distance & Constructive Upper Bound & Lower Bound & Gap \\ [0.5ex] 
\hline\hline
$n\!\times\!1$ & 2 & 1 & 1 & 0 \\
\hline
$n\!\times\!2$ & 3 & $\lceil\log_2(n+1)\rceil$ & $\lceil\log_2(n+1)\rceil$ & 0 \\
\hline
$n\!\times\!3$ & 4 & $\lceil\log_2(n+1)\rceil + 1$ & $\lceil\log_2(n+1)\rceil$ & $\lesssim 1$ \\
\hline
$n\!\times\!4$ & 5 & $\min\{2\lceil\log_2(n+5)\rceil + 1, 2\lceil\log_2(2n+1)\rceil\}$ & $2\lceil\log_2(n+1)\rceil-1$ & $\lesssim 4$ \\
\hline
$n\!\times\!2t$ & $2t+1$ & $t\lceil\log_2\left(nt+1\right)\rceil$ & $t\lceil\log_2(n)\rceil$ & $\lesssim t\log_2(t)$ \\
\hline
$n\!\times\!(2t+2)$ & $2t+2$ & $t\lceil\log_2\left(nt+2\right)\rceil+1$ & $t\lceil\log_2(n)\rceil+1$ & $\lesssim t\log_2(t)$ \\
\hline
\end{tabular}
\label{Summary-TE-Table}
\end{table*}

\subsection{\texorpdfstring{$e$}{e}-TE Codes Where \texorpdfstring{$e>L$}{e>L}}\label{sec:TE-codes-large-e}
Claim~\ref{clm:aux-for-const-TE-thm} holds only when $e\le L$. As of the writing of this paper, it is still an open problem whether Construction~\ref{construction:linear-TE-code} can be extended for the case where $e>L$, or not. The challenge becomes perceptible when one just examines the scenario in which $e=L+1$. In this case, the multiset $\cJ(\bm{\cH},\bm{p})$ for $\bm{p}$ such that $\|\bm{p}\|_1=L+1$, can have an entire row of $\bm{\cH}$ and also one other entry from the last column. Since we cannot know which entry it will be, every entry in the first column should differ from any other entry in $\bm{\cH}$, which results in an immediate decrease in terms of the optimality of the code.
However, there exist other constructions, where the assumption on $e$ is not required, and those will be discussed next.

First, the vector construction, where the $n\times L$ array is treated as an length-$nL$ vector. Therefore, the classical theory of erasure-correcting codes satisfies the model and no requirements such as $e\le L$ exists, but the construction does not use the special structure of the position of the erasures.
Next, we describe the Cartesian product construction as in~\cite{zhou2016bch}. The main idea here is to use independent erasure-correcting codes for every column. That is, the $j$-th column, $j\in [L]$, is a codeword in an $(n,M_j,\lceil d/(L-j+1)\rceil)$, where $d$ represents the merged code being a $(d-1)$-TE code. Also in this code, as in the vector construction, the correctness of the code does not rely on the condition of $e\le L$.
Moreover, the entire BCH codes that were constructed in~\cite{zhou2016bch} can correct erasures without restrictions on the number of erasures. Also, the construction from~\cite[Section 3.5]{raviv2020hierarchical} can correct at most $r$ erasures at the end of every row, which is $nr$ erasures in total, but with a restriction on the number of erasures in each row. The other condition for this code to exist, is $L\ge n \ge r$.

Next, we propose a construction of an $e$-TE code, in which the restriction that $e \leq L$ is removed. 
The results we achieved for $L=2,3,4$ and $e=2,3,4,5$, using this construction are summarized in Table~\ref{Summary-hasse-construction-Table}.
The bold entries represent cases in which Construction~\ref{construction:large-e-TE-code} is at least as good as the known state of the art results, in terms of the number of redundancy bits.

This construction is inspired by ideas from~\cite{PascalAshwin2007,zhou2016bch}, and is based on the \emph{Hasse derivative}, as defined in~\cite{Hasse1936} and is presented next. 
For any non-negative integer $i$, the $i$-th
Hasse derivative of a polynomial $f(x)=\sum_{k=0}^{d}a_{k}x^{k}\in\mathbb{F}_q[x]$ is
$f^{(i)}(x)\triangleq\sum_{k=0}^{d}\binom{k}{i}a_{k}x^{k-i}$,
where $\binom{k}{i}=0$ for all $k<i$.
Moreover, a fundamental property of the Hasse derivative~\cite[Lemma 6]{PascalAshwin2007} is that for every non-zero polynomial $f(x)\in\mathbb{F}_q[x]$, an element $\beta\in\mathbb{F}$ (where $\mathbb{F}$ is $\mathbb{F}_q$ or any extension field of $\mathbb{F}_q$), is a root of multiplicity $\ell$ if and only if $f^{(s)}(x)=0$ for $0 \le s < 
\ell$ and $f^{(\ell)}(x) \neq 0$.
The following is a construction which is based on the Hasse derivative.

\begin{construction} \label{construction:large-e-TE-code}
    Let $n,L,e$ be arbitrary positive integers, and let $m$ be the smallest integer such that $q=2^m>n$, that is, $m=\lceil\log_2(n+1)\rceil$. Let $\alpha\in\mathbb{F}_q$ be a primitive element. 
    Then, the TE parity check matrix of the code $\cC$ follows,
    \begin{align*}
        \bm{\cH}=
        \left[\begin{array}{ccccc}
            \bm{h}_{1,1} & \cdots & \bm{h}_{1,j} & \cdots & \bm{h}_{1,L} \\
            \bm{h}_{2,1} & \cdots & \bm{h}_{2,j} & \cdots & \bm{h}_{2,L} \\
            \vdots & \ddots & \vdots & \ddots &\vdots \\
            \bm{h}_{n,1} & \cdots & \bm{h}_{n,j} & \cdots & \bm{h}_{n,L} \\
        \end{array}\right]~,
    \end{align*}
    where, for $\beta_i=\alpha^{i}$, each entry is the vector
    \begin{align*}
        \bm{h}_{i,j}=
        \left[\begin{array}{c}
            \binom{0}{L-j}\beta_{i}^{0-L+j} \\
            \binom{1}{L-j}\beta_{i}^{1-L+j} \\
            \vdots \\
            \binom{e-1}{L-j}\beta_{i}^{e-1-L+j} \\
        \end{array}\right]\in\mathbb{F}_q^{e}~,i\in[n], j\in[L]~.
    \end{align*}
\end{construction}
\begin{remark}\label{remark:hasse-derivative-vector}
    The key observation here is that for each vector $\bm{h}_{i,j}$, it holds that for any vector $f=(f_0,f_1,\ldots,f_{e-1})\in\mathbb{F}_{q}^{e}$, which represents a polynomial $f(x)=\sum_{\ell=0}^{e-1}f_{\ell}x^{\ell}\in\mathbb{F}_{q}[x]$, the inner product $f\cdot\bm{h}_{i,j}$ is the $(L-j)$-th Hasse derivative of $f$, evaluated at $\beta_i$.
    Specifically, if there exists an index $j$, such that $f\cdot\bm{h}_{i,\ell}\!=\!0$ for $\ell \in \{L,L\!-\!1, \ldots, j\}$ and $f\cdot\bm{h}_{i,j-1}\!\neq\!0 $, then $\beta_i$ is a root of multiplicity $L\!-\!j\!+\!1$ of $f$.
\end{remark}

The following claim proves that Construction~\ref{construction:large-e-TE-code} is an $e$-TE code. 
Although the proof follows from the same ideas as in~\cite{PascalAshwin2007} we include it for completeness.

\begin{claim}\label{clm:constr-large-e-works}
    Let $\cC$ be the resulting code from Construction~\ref{construction:large-e-TE-code}. Then, $\rho_{\text{TE}}(\cC)=e+1$.
\end{claim}
\begin{IEEEproof}
    By using Claim~\ref{TE-min-dist-claim}, it needs to be proved that any multiset $\cJ(\bm{\cH},\bm{p})$ of cardinality $e$ is a linearly independent multiset, and therefore let $\cJ$ be an arbitrary multiset as such.
    Next, let $A\in\mathbb{F}_q^{e \times e}$ be the matrix that is obtained from placing all the vectors in $\cJ$ as columns of $A$.
    We want to show that $A$ has full rank. Therefore, assume, for the sake of contradiction, that there exists a nonzero vector $f\in\mathbb{F}_{q}^{e}$ such that $f\cdot A = 0$.
    Then, using the observation in Remark~\ref{remark:hasse-derivative-vector}, the polynomial $f(x)$ must have at least $e$ roots (including multiplicities), but it is of degree $e-1$, which is a contradiction.
    
\end{IEEEproof}

\begin{remark}
    We note the similarity between our approach and the Universally Decodable Matrices (UDM) construction introduced by~\cite{PascalAshwin2007}.
    In both cases, the properties of the Hasse derivative are applied to ensure the existence of a full-rank matrix, as demonstrated in Claim~\ref{clm:constr-large-e-works}.
    However, while \cite{PascalAshwin2007} applied this technique for encoding purposes and with different parameters, we have adapted it for our specific context.
    To maintain clarity and consistency within our paper, we presented Construction~\ref{construction:large-e-TE-code} using our established notation.
    Furthermore, as will be discussed in the rest of this subsection, we derived interesting results on the redundancy of Construction~\ref{construction:large-e-TE-code} for certain cases, an aspect not addressed in~\cite{PascalAshwin2007}.
\end{remark}

In order to analyze the number of redundancy bits that Construction~\ref{construction:large-e-TE-code} requires, one needs to expand the entries in every $\bm{h}_{i,j}$ into binary columns of length $m$, and then to compute the rank of the binary expansion of $\bm{\cH}$. This procedure is similar to the one done in analysis of alternant codes. Therefore, the actual number of redundancy bits depends on the specific parameters.

However, we provide next an example of a proof technique, for achieving a better upper bound for specific cases. The technique includes modifying the construction slightly by adding an additional row of parity. This will be demonstrated on a $5$-TE code of length $n\times 2$ that uses only $2(\lceil\log_2(n)\rceil+1)$ redundancy bits. This is better than the vector construction (by $1$ bit, as one can verify), which is the only construction that fits those parameters.

\begin{claim}\label{claim:hasse-construction-example}
    Let $n$ be a positive integer and $q=2^m$, where $m$ is the smallest integer such that $2^m > n$. Let $\alpha\in\mathbb{F}_q$ be a primitive element and $\beta_i=\alpha^i$. 
    \begin{align*}
        \bm{\cH}=
        \left[\begin{array}{cc}
            \bm{h}_{1,1} & \bm{h}_{1,2} \\
            \bm{h}_{2,1} & \bm{h}_{2,2} \\
            \vdots & \vdots \\
            \bm{h}_{n,1} & \bm{h}_{n,2} \\
        \end{array}\right]~,
    \end{align*}
    where
    \begin{align*}
        \bm{h}_{i,1}=
        \left[\begin{array}{cc}
            1 \\
            0 \\
            1 \\
            \beta_i^2  \\
        \end{array}\right]~,\quad
        \bm{h}_{i,2}=
        \left[\begin{array}{cc}
            0 \\
            1 \\
            \beta_i \\
            \beta_i^3  \\
        \end{array}\right]~.
    \end{align*}
    
    Then, $\bm{\cH}$ is a TE parity check matrix of a $5$-TE code.
\end{claim}
\begin{IEEEproof}
    First, let
    \begin{align*}
        \bm{\cH}^{\prime}=
        \left[\begin{array}{cc}
            \bm{h}_{1,1}^{\prime} & \bm{h}_{1,2}^{\prime} \\
            \bm{h}_{2,1}^{\prime} & \bm{h}_{2,2}^{\prime} \\
            \vdots & \vdots \\
            \bm{h}_{n,1}^{\prime} & \bm{h}_{n,2}^{\prime} \\
        \end{array}\right],
    \end{align*}
    where
    \begin{align*}
        \bm{h}_{i,1}^{\prime}=
        \left[\begin{array}{cc}
            1 \\
            0 \\
            1 \\
            0 \\
            \beta_i^2  \\
            0 \\
        \end{array}\right],~
        \bm{h}_{i,2}^{\prime}=
        \left[\begin{array}{cc}
            0 \\
            1 \\
            \beta_i \\
            \beta_i^2 \\
            \beta_i^3  \\
            \beta_i^4 \\
        \end{array}\right]~.
    \end{align*}
    
    Notice that an entry is added in the beginning of each $\bm{h}_{i,j}^{\prime}$, although from Claim~\ref{clm:constr-large-e-works} one can verify that $\bm{\cH}^{\prime}$ is a $5$-TE code.
    The way to prove this, is by showing that the removed rows ($4$-th and $6$-th) in every $\bm{h}_{i,j}^{\prime}$ can be recovered from the retained rows.
    Let $\bm{p}$ be a pattern of erasures and $\cJ(\bm{\cH}^{\prime},\bm{p})$ be the respective multiset of columns from $\bm{\cH}^{\prime}$. Suppose that $\cI=\{(i_1,j_1),(i_2,j_2),\ldots,(i_5,j_5)\}$ denotes the indices of those columns in $\cJ(\bm{\cH}^{\prime},\bm{p})$. 
    Also, let $\cI_{2}=\{i \colon (i,2)\in\cI\}$, that is all the indices that indicate erasures in the second (i.e. last) column, and $\cI_{1}=\{i \colon (i,1)\in\cI\}$, that is, the indices of erasures in the first (i.e., penultimate) column.
    
    Next, denote
    \begin{align*}
        R_{\ell}=\sum_{(i,j)\in\cI}c_{i,j}(\bm{h}_{i,j}^{\prime})_{\ell} = \sum_{i\in\cI_{1}}c_{i,1}(\bm{h}_{i,1}^{\prime})_{\ell}+\sum_{i\in\cI_{2}}c_{i,2}(\bm{h}_{i,2}^{\prime})_{\ell}~,
    \end{align*}
    
    where $c_{i,j}\in\mathbb{F}_2$, and notice that $\cI_{1}$ and $\cI_{2}$ are not disjoint, which is desired.
    
    Next, we prove that $R_{4}=(R_{3})^2+R_1$, so that $R_{4}$ can be removed from the parity check matrix, using the characteristic of $\mathbb{F}_q$, which is $2$, as follows,
    \begin{dmath*}
        R_{4}=\sum_{i\in\cI_{1}}c_{i, 1}(\bm{h}_{i,1}^{\prime})_{4}+\sum_{i\in\cI_{2}}c_{i,2}(\bm{h}_{i,2}^{\prime})_{4}=\sum_{i\in\cI_{2}}c_{i,2}\beta_i^2 \\
        =\sum_{i\in\cI_{2}}c_{i,2}\beta_i^2 + \sum_{i\in\cI_{1}}c_{i,1}\cdot1 + \sum_{i\in\cI_{1}}c_{i,1}\cdot1 \\
        =\left(\sum_{i\in\cI_{2}}c_{i, 2}\beta_i + \sum_{i\in\cI_{1}}c_{i,1}\cdot1\right)^2 + \sum_{i\in\cI_{1}}c_{i,1}\cdot1 \\
        =(R_{3})^2+R_1~.
    \end{dmath*}
    Noting that also $R_6=(R_4)^2$ completes the proof, since by removing $R_4,R_6$, the matrix $\bm{\cH}^{\prime}$ becomes the desired $\bm{\cH}$.
    
\end{IEEEproof}

The same ideas from the proof of the previous claim can be applied to a few additional parameter regimes, which are highlighted in Table~\ref{Summary-hasse-construction-Table}, that achieve the best known results.

\begin{table}[htb]
\centering
\caption{Upper Bounds On The Redundancy Bits Of Construction~\ref{construction:large-e-TE-code}.}
\begin{tabular}{| c | c | c |}
\hline
& $L=2$ & $L=3,4$  \\ [0.5ex] 
\hline
$e=2$ & $\bm{\log_{2}(n)+1}$ & $\bm{\log_{2}(n)+1}$ \\
\hline
$e=3$ & $\bm{\log_{2}(n)+2}$ & $\log_{2}(n)+3$ \\
\hline
$e=4,5$ & $\boldsymbol{2\log_{2}(n)+2}$ & $\boldsymbol{2\log_{2}(n)+3}$ \\
\hline
\end{tabular}
\label{Summary-hasse-construction-Table}
\end{table}

\section{\texorpdfstring{$(t,s)$}{(t,s)}-Deletion-Correcting Codes}
\label{sec:(t,s)-del-corr}
In the process of constructing optimal $(t,s)$-DC codes, we first introduce a construction of a $(t,1)$-DC code, and an explicit encoder for this construction will be presented afterwards. Then, a generalization for an encoder of a $(t,s)$-DC code will conclude this Section.
In Section \ref{sec:Bounds:DC Upper Bounds}, it is shown that, in some cases, those constructions are optimal.

\subsection{\texorpdfstring{$(t,1)$}{(t,1)}-DC Codes}\label{subsec:(t,1)-DC-const}
In order to correct a single deletion in a length-$n$ vector, the Varshamov-Tenengolts (VT) codes~\cite{VT1965} are the common solution and will also form the basis of the following construction. 
For $0\leq a\leq L$, denote $\text{VT}_{a}(L)$, as defined in \cite{VT1965},
\begin{align*}
    \text{VT}_{a}(L) = \left\{\bm{x}\in \{0,1\}^L \colon \sum_{j=1}^{L} j x_{j} \equiv a \Mod{L+1} \right\}~.
\end{align*}

Next, we construct a code based on the tensor product code concept introduced in~\cite{Wolf06}. 
To facilitate this construction, we will first define a variant of VT codes.
Denote $h = \lceil \log_2(L+1)\rceil$, such that $\mathbb{F}_{2^h}$, the field of $2^h$ elements, is the smallest extension field of $\mathbb{F}_{2}$ that has at least $L+1$ elements.
For each row $\bm{x}_i = (x_{i,1},\dots,x_{i,L})$, define the \emph{$q$-syndrome} to be $s_{q}(\bm{x}_i) \equiv \sum_{j=1}^{L} j x_{i,j}\Mod{q}$.
Let $\phi:\mathbb{Z}_{2^h} \to \mathbb{F}_{2^h}$ be a bijection and
denote $\phi(s_{2^h}(\bm{x}_i))$ as $\sigma(\bm{x}_i)$, which is the syndrome of $\bm{x}_i$ as an element of the field $\mathbb{F}_{2^h}$.

A definition for a variation of the binary VT codes of length $L$ follows, 
$$\text{VT}^{2^h}_{a}(L) = \left\{\bm{x}\in\mathbb{F}_{2}^{L} \colon \sum_{j=1}^{L} j x_{j}\equiv a \Mod{2^h} \right\}~,$$
where now it holds that $0\leq a\leq 2^h-1$. One can verify that $\text{VT}^{2^h}_{a}(L)$ can correct a single deletion using the same decoding algorithm as the original VT codes~\cite{Levenshtein_SPD66}.
The following is an implicit construction of a $(t,1)$-DC code.

\begin{construction}\label{const:(t,1)-implicit-DC}
Let $\cC$ be a code of length $n$, that can correct $t$ erasures over $\mathbb{F}_{2^h}$. Then, 
\begin{align*}
    \cC_{\text{DC}}(n,L,t,\cC) \coloneqq \big\{ \bm{X}=\left(\bm{x}_1,\dots,\bm{x}_n\right)\in\mathbb{F}_2^{n\times L} \colon \\ 
    \left( \sigma \left(\bm{x}_1\right), \dots, \sigma(\bm{x}_n) \right) \in \cC \big\}~.
\end{align*}
\end{construction}
As discussed before, Construction~\ref{const:(t,1)-implicit-DC} yields a tensor product code that relies on the syndromes of the variant of the VT code, such that the possible values of the syndrome can be treated as field elements, and therefore the $n$ syndromes comprise the symbols from a codeword of the non-binary code $\cC$ over the same field.

\begin{remark}\label{remark:better-codes-over-ring}
    Note that increasing the parameter $h$ results in codes with larger redundancy.
    One can choose to work with the original VT code, and then construct the non-binary code over the ring $\mathbb{Z}_{L+1}$, instead of a field, if there exists an optimal code over this ring with the required parameters.
\end{remark}

Next, we prove that Construction~\ref{const:(t,1)-implicit-DC} is a $(t,1)$-DC code.

\begin{lemma}\label{lemma:const:(t,1)-implicit-DC}
    The code $\cC_{\text{DC}}(n,L,t,\cC)$ is a $(t,1)$-DC code.
\end{lemma}
\begin{IEEEproof}
    The proof is based on the idea that due to the deletions, each erroneous row is shorter, and therefore the location (index) of these rows is known.
    Assume that $s\le t$ rows suffer from one deletion each and denote its indices as $i_1,\dots,i_s$. Since $n-s$ rows are error-free, their syndromes can be computed and mapped into $\sigma(\bm{x}_j)$, where $j\in[n]\setminus\{i_1,\dots,i_s\}$.
    Now $\cC$ can recover $\sigma(\bm{x}_{i_1}),\dots,\sigma(\bm{x}_{i_s})$ such that $\left(\sigma\left(\bm{x}_1\right),\dots,\sigma(\bm{x}_n)\right)$ can be used truthfully as syndromes for each of the $\text{VT}^{2^h}_{\sigma(\bm{x}_i)}(L)$ codewords, which is the $i$-th row.
    Therefore, the rows $\bm{x}_{i_1},\dots,\bm{x}_{i_s}$ can be recovered.
\end{IEEEproof}
\vspace{5pt}
Before we present an explicit encoder for Construction \ref{const:(t,1)-implicit-DC}, we give an upper bound on the minimal number of redundancy bits that are required to construct a $(t,1)$-DC code, based on Construction \ref{const:(t,1)-implicit-DC}.

\begin{theorem}\label{thm:redundancy-of-(t,1)-implicit-DC}
    Let $R_{\text{Opt}}$ be the number of redundancy symbols of $\cC_{\text{Opt}}$, a linear code of length $n$, that can corrects $t$ erasures over $\mathbb{F}_{2^h}$. 
    Then, there exists a code $\cC_{\text{DC}}(n,L,t,\cC_{\text{Opt}})$ with a redundancy of at most $R_{\text{Opt}}\cdot h = R_{\text{Opt}}\cdot\left\lceil \log_2(L+1)\right\rceil$ bits.
\end{theorem}
\begin{IEEEproof}
    Note that any coset of $\cC_{\text{Opt}}$ is a distinct $t$-erasures-correcting code, and there are $q^{R_{\text{Opt}}}$ cosets, where $q=2^{h}$.
    Thus, there are $q^{R_{\text{Opt}}}$ distinct $\cC_{\text{DC}}(n,L,t,\cdot)$ codes, one for each coset, and all those distinct $\cC_{\text{DC}}(n,L,t,\cdot)$ codes create a partition of the space of all the $2^{nL}$ binary arrays of size $n\times L$.
    Therefore, using the pigeonhole principle, there exists at least one code with $2^{nL}/q^{R_{\text{Opt}}}=2^{nL-R_{\text{Opt}}\cdot\left\lceil \log_2(L+1)\right\rceil}$ codewords.
\end{IEEEproof}

Note that it implies the existence of a $(1,1)$-DC code with only $\log_2(L+1)$ bits of redundancy.
Also, if $\cC_{\text{Opt}}$ is an MDS code, the upper bound on the redundancy is exactly $t\lceil\log_2(L+1)\rceil$ bits, and by looking at the discussion after Theorem \ref{thm:summary-of-DC-bounds}, one can verify optimality, assuming the VT codes are optimal $1$-deletion-correcting codes.

\begin{corollary}\label{cor:redundancy-of-(t,1)-DC-codes}
    There exists a $\cC_{\text{DC}}(n,L,t,\cC_{\text{Opt}})$ code with a redundancy of at most $t(\log_2(n)+\lceil\log_2(L+1)\rceil)$ bits.
\end{corollary}
\begin{IEEEproof}
    By using alternant code over $\mathbb{F}_{2^h}$ as in~\cite[Section 5.5]{RothBook2006}, one has a code $\cC$ with a redundancy of at most $t\cdot m$ symbols, where $m$ is an integer such that $2^m\ge (2^{h})^n$.
    
    Therefore, an optimal choosing of $m$ is $\lceil\log_{2^{h}}(n)\rceil$.
    Finally, using the same proof as in Theorem~\ref{thm:redundancy-of-(t,1)-implicit-DC}, where $R \leq t\cdot\lceil\log_{2^{h}}(n)\rceil$ and thus there exists a code with a redundancy of at most 
    \begin{align*}
        R\cdot h &\leq t\cdot\lceil\log_{2^{h}}(n)\rceil\cdot h
        = t\cdot\left\lceil\frac{\log_{2}(n)}{h}\right\rceil\cdot h
        \leq t(\log_{2}(n)+h)~.
    \end{align*}
\end{IEEEproof}

\begin{remark}
    Note that if $n=c\cdot2^{h}$ for some positive integer $c$, then the bound in Corollary~\ref{cor:redundancy-of-(t,1)-DC-codes} is $t(\log_{2}(n))$.
\end{remark}

Finally, an explicit encoder of Construction~\ref{const:(t,1)-implicit-DC} is presented.
In~\cite{Abdel-Ghaffar98}, a systematic encoder of $\text{VT}_{a}(L)$, where the redundancy bits are in the $\{2^{i}\}_{i=0}^{h-1}$ coordinates was introduced. 
Then, an encoder for $\text{VT}^{2^h}_{a}(L)$, denoted as $\cE_{\text{VT}}^{(L,a)}$, can be obtained using the same idea, since one can verify that it requires exactly the same entries, $\{2^{i}\}_{i=0}^{h-1}$, for the encoding process.
$\cE_{\text{VT}}^{(L,a)}$ is defined as follows.
First, it receives a binary input vector $\bm{d}$ of length $L-h$, then outputs a codeword $\bm{x}$ of length $L$, such that $\cE_{\text{VT}}^{(L,a)}(\bm{d})=\bm{x}\in \text{VT}^{2^h}_{a}(L)$.

\begin{encoder}\label{enc:(t,1)-DC}
Let $\text{E}\colon {(\mathbb{F}_{2^{h}})}^{n-R} \to {(\mathbb{F}_{2^{h}})}^{n}$ be a systematic encoder of a linear code $\cC$ of length $n$, that can correct $t$ erasures over $\mathbb{F}_{2^{h}}$. Lastly, denote $K\!=\!nL\!-\!Rh$.

\textbf{Input:} $\bm{D}=(d_1,\dots,d_{K})\in\mathbb{F}_2^K$.

\textbf{Output:} $\bm{X}=\left(\bm{x}_1,\dots,\bm{x}_n\right)\in\mathbb{F}_2^{n\times L}$.
\begin{enumerate}
    \item For $i\in[n-R], j\in[L] \colon x_{i,j}\leftarrow d_{(i-1)L+j}$.
    \item Compute $\{\sigma(\bm{x}_1),\dots,\sigma(\bm{x}_{n-R})\}$. 
    Then compute $\text{E}\left(\sigma(\bm{x}_1),\dots,\sigma(\bm{x}_{n-R})\right)$ and denote the output as $(\sigma(\bm{x}_1),\dots,\sigma(\bm{x}_{n}))$.
    \item Take the remaining $R(L-h)$ bits of $\bm{D}$ and split them into $R$ vectors of $L-h$ bits, denoted by $\{\bm{d}_1,\dots,\bm{d}_R\}$.
    \item For $i\in[R]\colon a_{n\!-R\!+i}\leftarrow\phi^{-1}\left(\sigma(\bm{x}_{n\!-R\!+i})\right)$, that is recovering from $\sigma(\cdot)$ the expected syndrome value of the $(n-R+i)$-th row, which is denoted as $a_{n-R+i}$.
    \item For $i\in[R]\colon \bm{x}_{n-R+i}\leftarrow\cE_{\text{VT}}^{(L,a_{n\!-R\!+i})}(\bm{d}_i)$, i.e., encode the VT codeword $\bm{x}_{n-R+i}$ from $\bm{d}_i$ as defined in step 3, based on the given syndrome value $a_{n\!-R\!+i}$.
    \item Return $\bm{X}$.
\end{enumerate}
\end{encoder}
The decoding steps are straightforward, using the decoders of $\cC$ and $\text{VT}^{2^h}_{a}(L)$ in the same way as in the proof of Lemma~\ref{lemma:const:(t,1)-implicit-DC}, which proves the correctness, since one can readily verify that the output of Encoder~\ref{enc:(t,1)-DC} is a codeword of $\cC_{\text{DC}}(n,L,t)$ from Construction \ref{const:(t,1)-implicit-DC}.
Moreover, as one can verify, Encoder~\ref{enc:(t,1)-DC} uses exactly $R_{\text{Opt}}\cdot\left\lceil \log_2(L+1)\right\rceil$ bits of redundancy, which is an explicit construction for the existance proof in Theorem~\ref{thm:redundancy-of-(t,1)-implicit-DC}.
Finally, we give an example of Encoder~\ref{enc:(t,1)-DC}, for constructing a $(2,1)$-DC code over the space of $7\times 5$ binary arrays.

\begin{example}\label{ex:(t,1)-DC-enc}
In Figure~\ref{fig:(t,1)-DC-enc-example}, observe an example of an output of Encoder~\ref{enc:(t,1)-DC}.  
It receives $29$ bits as an input, denoted as $\bm{D}=(x_{1,1}, \ldots, x_{5,5}, x_{6,3}, x_{6,5}, x_{7,3}, x_{7,5})\in\mathbb{F}_2^{29}$ and are shown with the white background in the $7\times 5$ array codeword. 
The encoder computes the syndromes $(\sigma(\bm{x}_1),\ldots,\sigma(\bm{x}_5))$ and then, using an encoder of a $[7,5,3]$ code over $\mathbb{F}_{2^3}$, computes also 
$\sigma(\bm{x}_6)$ and $\sigma(\bm{x}_7)$. The syndromes are presented with dashed lines on the right side of the codeword. 
Lastly, using the systematic encoder for the VT codes, it computes the redundancy values in the last $2$ rows, marked with gray background.
Therefore, it is an example of size $(7\times 5)$ binary array code with $6$ bits of redundancy.
\end{example}

\begin{figure}
\centering
\includegraphics[width=0.75\linewidth]{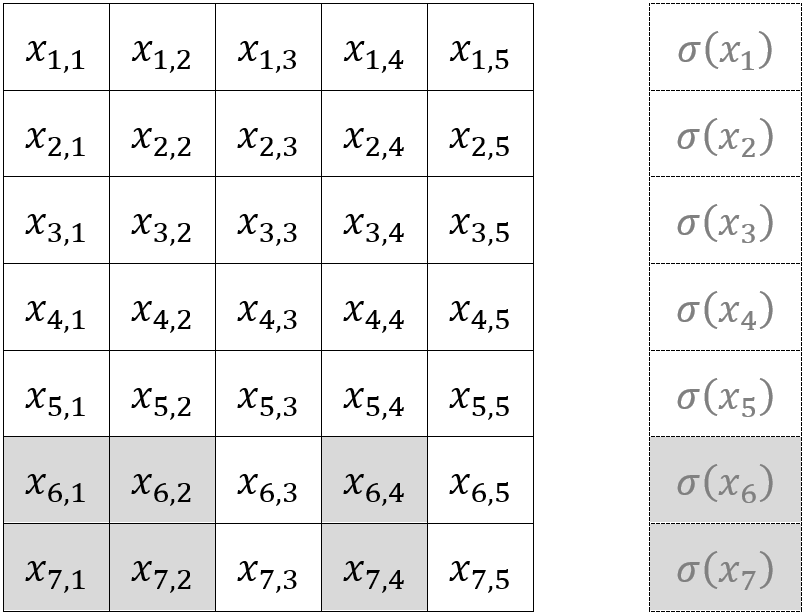}
\caption{Example of an output of Encoder~\ref{enc:(t,1)-DC} for a $(2,1)$-DC code over binary arrays of length $7\times 5$. Illustration of Example~\ref{ex:(t,1)-DC-enc}.}
\label{fig:(t,1)-DC-enc-example}
\end{figure}

\subsection{Construction of a \texorpdfstring{$(t,s)$}{(t,s)}-DC Code}\label{subsec:(t,s)-DC-enc}
This section is based on~\cite{Sima2020}.
First, denote by $\cC_s$ the authors' $s$-deletion-correcting code over the vector space $\mathbb{F}_{2}^{L+r}$, which is composed from all the codewords $\left(\bm{c},g(\bm{c})\right)$, where $\bm{c}\in\mathbb{F}_{2}^{L}$, and $g(\bm{c})$\footnote{For shortening, we use $g$ instead of the modular version in~\cite{Sima2020}, which is denoted by $g_c$} is used by the authors as an $s$-deletion-correcting hash for $\bm{c}$ of size $r\!=\!4s\log_2(L)\!+\!o(\log_2(L))$.
The existence of such a code is proved in~\cite{Sima2020}.
Finally, denote by $\Gamma(\bm{c})$ the mapping of  $g(\bm{c})$ to an element of $\mathbb{F}_{2^{r}}$.

\begin{construction}\label{const:(t,s)-implicit-DC}
Let $n$, $L$, $t$ and $s$ be positive integers, and let $r$ be the constant described previously. Next, let $\cC$ be a code of length $n$, that can correct $t$ erasures over $\mathbb{F}_{2^{r}}$. Then, 
\begin{align*}
    \cC_{s\text{-DC}}(n,L,t,\cC) \coloneqq \big\{\bm{X}=\left(\bm{x}_1,\dots,\bm{x}_n\right)\in\mathbb{F}_2^{n\times L} \colon \\ 
    \left( \Gamma\left(\bm{x}_1\right), \dots, \Gamma\left(\bm{x}_n\right) \right)\in \cC \big\}~.
\end{align*}
\end{construction}

The next lemma follows a proof outline similar to that of Lemma~\ref{lemma:const:(t,1)-implicit-DC}.
\begin{lemma}\label{lemma:const:(t,s)-implicit-DC}
    The code $\cC_{s\text{-DC}}(n,L,t,\cC)$ is a $(t,s)$-DC code.
\end{lemma}
\begin{IEEEproof}
    As in Lemma~\ref{lemma:const:(t,1)-implicit-DC}, due to the deletions, each erroneous row is shorter, and therefore the index of these rows is known.
    Assume that $w\le t$ rows suffer from one deletion each and denote its indices as $i_1,\dots,i_w$. Since $n-w$ rows are error-free, the value of their $g$ function can be computed and mapped into $\Gamma(\bm{x}_j)$, where $j\in[n]\setminus\{i_1,\dots,i_w\}$.
    Now $\cC$ can recover $\Gamma(\bm{x}_{i_1}),\dots,\Gamma(\bm{x}_{i_s})$ such that $\left(\Gamma\left(\bm{x}_1\right),\dots,\Gamma(\bm{x}_n)\right)$ can be mapped back to valid values of $\left(g\left(\bm{x}_1\right),\dots,g(\bm{x}_n)\right)$, and then the rows $\bm{x}_{i_1},\dots,\bm{x}_{i_w}$ can be recovered using the decoder from~\cite{Sima2020}, since each of these rows suffers at most $s$ deletions, and therefore the set of words $\{(\bm{x}_{i_j}, g(\bm{x}_{i_j}))\}_{j=1}^{w}$ also, which is the input to the decoder in~\cite{Sima2020}.

\end{IEEEproof}

Moreover, the next theorem is also using similar techniques as in Theorem~\ref{thm:redundancy-of-(t,1)-implicit-DC}.

\begin{theorem}\label{thm:redundancy-of-(t,s)-implicit-DC}
    Let $R_{\text{Opt}}$ be the number of redundancy symbols of $\cC_{\text{Opt}}$, a linear code of length $n$, that can corrects $t$ erasures over $\mathbb{F}_{2^{r}}$. 
    Then, there exists a code $\cC_{s\text{-DC}}(n,L,t,\cC)$ with a redundancy of at most $R_{\text{Opt}}\cdot r \approx R_{\text{Opt}} \cdot 4s\log_2(L)$ bits.
\end{theorem}
\begin{IEEEproof}
    Note that any coset of $\cC_{\text{Opt}}$ is a distinct $t$-erasures-correcting code, and there are $q^{R_{\text{Opt}}}$ cosets, where $q=2^{r}$.
    Thus, there are $q^{R_{\text{Opt}}}$ distinct $\cC_{s\text{-DC}}(n,L,t,\cdot)$ codes, one for each coset, and all those distinct $\cC_{s\text{-DC}}(n,L,t,\cdot)$ codes create a partition of the space of all the $2^{nL}$ binary arrays of size $n\times L$.
    Therefore, using the pigeonhole principle, there exists at least one code with $2^{nL}/q^{R_{\text{Opt}}}=2^{nL-R_{\text{Opt}}\cdot r}$ codewords.
    
\end{IEEEproof}

\begin{remark}
    We note that Construction~\ref{const:(t,s)-implicit-DC} and Theorem~\ref{thm:redundancy-of-(t,s)-implicit-DC} provide only an existential proof of $(t,s)$-DC codes. Since one cannot ensure that $\Gamma$ is surjective, an explicit encoder of Construction~\ref{const:(t,s)-implicit-DC} remains a more complex task, which is left future research.
\end{remark}

\renewcommand{\arraystretch}{1.3}
\begin{table*}[t]
\centering
\caption{Bounds on the redundancy of $(t,s)$-DC codes, where $h = \lceil \log_2(L+1)\rceil$ and $c>0$ is an integer.}
\begin{tabular}{||c c c c||}
\hline
Model & Restriction & Constructive Upper Bound & Asymptotic Lower Bound 
\\ [0.5ex] 
\hline\hline
$(t,1)$ & $n \leq 2^{h}+1$ & $th$ & $t\lceil\log_2(L)\rceil$ 
\\
\hline
$(t,1)$ & $n=c\cdot2^{h}$ & $t\log_2(n)$ & $\max\{th,\lfloor t/2 \rfloor \log_2(n)\}$ 
\\
\hline
$(t,1)$ & - & $t(\log_2(n)+h)$ & $\max\{th,\lfloor t/2 \rfloor \log_2(n)\}$ 
\\
\hline
$(t,s)$ & - & $t(\log_2(n)+4s\log_2(L) + o(\log_2(L))$ & $\lfloor t/2 \rfloor (\log_2(n) + \lfloor s/2 \rfloor \log_2(L))$ 
\\
\hline
\end{tabular}
\label{Summary-DC-Table}
\end{table*}

\section{\texorpdfstring{$(t,1,e)$}{(t,1,e)}-TED-Correcting Code}
\label{sec:(t,1,e)-TED-corr}
In this section we present a method of constructing $(t,1,e)$-TED-correcting code.
This construction and its encoder use a non-binary code that is based upon the following two pieces of information:
(i) The VT syndrome for each row, (ii) The value of the last $e$ bits of each row.

As in Section~\ref{subsec:(t,1)-DC-const}, let $h = \lceil \log_2(L+1)\rceil$, and for any array $\bm{X}=\left(\bm{x}_1,\dots,\bm{x}_n\right)\in\mathbb{F}_2^{n\times L}$ denote by $s_i$ the $2^h$-syndrome of the $i$-th row of $\bm{X}$ as in Section~\ref{subsec:(t,1)-DC-const}, which is to say that $s_i=\sum_{j=1}^{L}jx_{i,j}\Mod{2^h}$.
Moreover, denote the last $e$ bits of the $i$-th row of $\bm{X}$ by $\bm{x}_i^{(:e)}\triangleq(\bm{x}_i)_{[L-e+1\colon L]}$.
Next, for each row of $\bm{X}$, define the tuple $\left(s_i,\bm{x}_i^{(:e)}\right)$. 
Finally, let $q = 2^{h} \cdot 2^{e}=2^{h+e}$ and denote by $\theta_{\bm{x}_i}$ the mapping of the tuple $\left(s_i,\bm{x}_i^{(:e)}\right)$ to an element of $\mathbb{F}_q$.

\begin{construction}\label{cnstrct:TED}
    Let $\cC$ be an $[n,k,t\!+\!e\!+\!1]$ code over $\mathbb{F}_q$.
    Then,
    \begin{align*}
        \cC_{\text{TED}}(n,L,t,e) \coloneqq \big\{ \bm{X}\in\mathbb{F}_2^{n\times L} \colon \left(\theta_{\bm{x}_1},\dots,\theta_{\bm{x}_n}\right)\in \cC \big\}~.
    \end{align*}
\end{construction}
Next, is a proof for the correctness of Construction~\ref{cnstrct:TED}.

\begin{lemma}
    The code $\cC_{\text{TED}}(n,L,t,e)$ is a $(t,1,e)$-TED code.
\end{lemma}
\begin{IEEEproof}
    A codeword $\bm{X}\in\cC_{\text{TED}}(n,L,t,e)$ is received with at most $t+e$ deletions, some of which are arbitrary deletions and some are tail-erasures.
    Therefore, there are at most $t+e$ rows for which one cannot compute the tuple $\left(s_i,\bm{x}_i^{(:e)}\right)$, and since $\cC$ can correct $t+e$ erasures, the entire codeword $\left(\theta_{\bm{x}_1},\dots,\theta_{\bm{x}_n}\right)\in \cC$ can be recovered.
    Next, one can recover the codeword $\bm{X}$ row by row.
    For each row, if it is of length $L$, it does not suffer from any deletion.
    If it is of length $L-1$, it can be recovered using the VT syndrome, regardless of the source of the deletion (arbitrary or TE).
    Finally, if it is of length $L-k$ for some $1<k\le e+1$, then it suffered from at most $1$ arbitrary deletion, and the rest are $k-1$ TEs, or just $k$ TEs. Thus, one can recover the last $k-1$ bits, as it is saved in the tuple $\left(s_i,\bm{x}_i^{(:e)}\right)$, and treat the last deleted bit as an arbitrary deletion, regardless of the source.
    
\end{IEEEproof}

Similarly to Remark~\ref{remark:better-codes-over-ring}, one can choose to work with the original VT codes, and then $q=2^{e}\cdot(L+1)$ is not a prime power, and afterwards construct the non-binary code over the ring $\mathbb{Z}_{q}$, instead of a field.
Moreover, as a proof of existence of codes that are built according to Construction~\ref{cnstrct:TED}, an explicit encoder is provided next.

\begin{encoder}\label{enc:(s,1,e)-TED-nonbinary-code}
Let $\text{E}\colon {(\mathbb{F}_{q})}^{n-R} \to {(\mathbb{F}_{q})}^{n}$ be a systematic encoder of an optimal linear code $\cC$ of length $n$, that can correct $t+e$ erasures over $\mathbb{F}_{q}$. Also, let $\cE_{\text{VT}}^{(L,a)}$ be a systematic VT code encoder, as defined in Section~\ref{sec:(t,s)-del-corr}. Lastly, denote $K_{\text{TED}}=nL-R(e+h)$.

\noindent\textbf{Input:} $\bm{D_{\text{TED}}}=(d_1,\dots,d_{K_{\text{TED}}})\in\mathbb{F}_2^{K_{\text{TED}}}$.

\noindent\textbf{Output:} $\bm{X}=\left(\bm{x}_1,\dots,\bm{x}_n\right)\in\mathbb{F}_2^{n\times L}$.
\begin{enumerate}
    \item For $i\in[n-R], j\in[L] \colon x_{i,j}\leftarrow d_{(i-1)L+j}$, and denote $\bm{x}_i^{(:e)}=(x_{i,L-e+1},\ldots,x_{i,L})$.
    \item Compute $\{s_1,\dots,s_{n-R}\}$, and then $\{\theta_{\bm{x}_1},\dots,\theta_{\bm{x}_{n-R}}\}$.
    \item Compute $\text{E}\left(\theta_{\bm{x}_1},\dots,\theta_{\bm{x}_{n-R}}\right)$ and denote the output as $\left(\theta_{\bm{x}_1},\dots,\theta_{\bm{x}_{n}}\right)$.
    \item Take the remaining $R(L-e-h)$ bits of $\bm{D}$ and split them into $R$ vectors of length $L-e-h$, denoted by $\{\bm{d}_1,\dots,\bm{d}_{R}\}$.
    \item For $i\in\{n-R+1,\ldots,n\}$, extract the tuple $\left(s_i,\bm{x}_i^{(:e)}\right)$, computed as $\theta_{\bm{x}_i}$ in Step 3.
    \item Let ${\widetilde{\bm{d}}}_i$ be the appending of $\bm{x}_i^{(:e)}$ to the end of $\bm{d}_i$.
    \item Compute $\bm{x}_{i}=\cE_{\text{VT}}^{(L,s_{i})}({\widetilde{\bm{d}}}_i)$, i.e. encode the VT codeword $\bm{x}_{i}$ from ${\widetilde{\bm{d}}}_i$, such that the last $e$ bits of $\bm{x}_{i}$ are $\bm{x}_i^{(:e)}$ and the syndrome value of $\bm{x}_{i}$ is $s_{i}$.
    \item Return $\bm{X}$.
\end{enumerate}
\end{encoder}

Note that Encoder~\ref{enc:(s,1,e)-TED-nonbinary-code} requires that $e<(L+1)-2^{h-1}$.
Also, it can be verified that Encoder~\ref{enc:(s,1,e)-TED-nonbinary-code} outputs a codeword of $\cC_{\text{TED}}(n,L,t,e)$ from Construction~\ref{cnstrct:TED}.
\begin{corollary}\label{cor:TED-results}
The following are results regarding the redundancy of $\cC_{\text{TED}}(n,L,t,e)$, denoted as $R_{\text{TED}}$.
    \begin{enumerate}
        \item The redundancy of the code $\cC_{\text{TED}}(n,L,t,e)$, using Encoder~\ref{enc:(s,1,e)-TED-nonbinary-code} is $nL-K_{\text{TED}}=R(e+h)$.
        \item Using an alternant code as $\cC$ in Encoder~\ref{enc:(s,1,e)-TED-nonbinary-code} implies
        $$R_{\text{TED}} \le \left(e+h\right)\left(\frac{t+e}{2}\right)\log_2(n)~.$$
        \item If $n\le 2^{h+e}$, using an MDS code code as $\cC$ in Encoder~\ref{enc:(s,1,e)-TED-nonbinary-code} implies
        $$R_{\text{TED}} \le \left(e+\lceil \log_2(L+1)\rceil\right)\left(t+e\right)~.$$
        \item If $n\le 2^{h+1}$, using an MDS code code as $\cC$ in Encoder~\ref{enc:(s,1,e)-TED-nonbinary-code} implies a $(1,1,1)$-TED code with $2+\lceil 2\log_2(L+1)\rceil$ bits of redundancy.
    \end{enumerate}
\end{corollary}

Note that the forth result in Corollary~\ref{cor:TED-results} is asymptotically optimal, when compared to the result in Theorem~\ref{thm:bound-TED}.
Finally, note that one can use a $(t+e)$-deletion-correcting code over vectors of length $nL$ as in~\cite{Sima2020}, with a redundancy of 
$$4(t+e)(\log_2(nL))+o(\log(nL))~,$$
which in cases where $n>2^{h+e}$, can achieve better results in terms of redundancy than Encoder~\ref{enc:(s,1,e)-TED-nonbinary-code}.

\section{Upper Bounds on the Code Cardinalities}
\label{sec:Bounds}
In this section we consider upper bounds on the maximal cardinality of TE codes, $(t,s)$-DC codes, and $(1,1,1)$-TED codes. These bounds are then compared against the constructions presented in previous sections.

\subsection{Upper Bounds on the Size of Tail-Erasure Codes}
\label{sec:Bounds:TE Upper Bounds}
Let $\cA_{\text{TE}}(n,d)$ denote the maximal cardinality of a $(d-1)$-TE  code over the set of $n\times (d-1)$ binary arrays.
Let $A(n,d)$ denote the maximal cardinality of a binary vector code of length $n$ with minimum Hamming distance $d$. 
The next observation follows, by noting that if the first $n-1$ columns of two codewords in a $(d-1)$-TE code are correspondingly equal, their last column must have Hamming distance at least $d$.

\begin{claim}\label{clm:first-TE-bound}
$\cA_{\text{TE}}(n,d) \leq A(n,d) \cdot 2^{n(d-2)}$.
\end{claim}
\begin{IEEEproof} To prove the result, suppose $\cC_{TE} \subseteq \mathbb{F}_{2}^{n \times (d-1)}$ is an $(d-1)$-TE code of maximal size. 
Next, decompose the codewords in $\cC_{TE}$ into two parts. 
Let $\cX \subseteq \mathbb{F}_2^{n \times (d-2)}$ be equal to the set of arrays that result by removing the last column from each of the codeword arrays from $\cC_{TE}$. 
For any $\textbf{Y} \in \cX$, let $S_{\textbf{Y}} = \left\{ \bfv \in \mathbb{F}_2^n : \begin{bmatrix} \textbf{Y} & \bfv^T \end{bmatrix} \in \cC_{TE} \right \}$.
In other words, $S_{\textbf{Y}}$ is the set of vectors that can be appended to $\textbf{Y}$ as the last column to generate a codeword. Since $\cC_{TE}$ can correct $d-1$ TEs, that can all occur in the last column, it follows that the minimum Hamming distance of the set of vectors in $S_{\textbf{Y}}$ is at least $d$, which implies that for any $\textbf{Y}\!\in\!\cX$, we have $\left| S_{\textbf{Y}} \right|\!\leq\! A(n,d)$.
One can verify that $|\cX| \leq 2^{n(d-2)}$, and it follows that,
\begin{align*}
   \left| \cC_{TE} \right| = \sum_{\textbf{Y} \in \cX} | S_{\textbf{Y}} | \leq \sum_{\textbf{Y} \in \cX} A(n,d) \leq A(n,d) \cdot 2^{n(d-2)},
\end{align*}
which completes the proof.

\end{IEEEproof}

Using the notations of Section \ref{sec:TE codes:Construction}, the above claim provides a simple lower bound on $R_{\text{TE}}(n,d)$, which is $$R_{\text{TE}}(n,d)\ge R(n,d) \ge \left\lceil\frac{d-1}{2}\right\rceil\log_2(n),$$ where the second inequality is derived from the sphere packing bound for vector codes of length $n$.

Next, a sphere packing bound is presented, which is tighter, but it is also harder to compute for larger values of minimum $\rho_{\text{TE}}$-distance of a TE code.
\begin{lemma}\label{sphere-packing-bound}
    The volume of a ball of radius $r$ in $\mathbb{F}_2^{n\times L}$, under the $\rho_{\text{TE}}$-metric, denoted as $V_{\text{TE}}(r)$, is given by
    \begin{align*}
            V_{\text{TE}}(r) = \sum_{k=0}^r\,\sum\limits_{\substack{t_1+2t_2+\dots\\+k^{\prime}t_{k^{\prime}}=k}}\left({\binom{n}{t_0,t_1,t_2,\dots,t_{k^{\prime}}}\cdot \prod_{i=1}^k 2^{(i-1)t_i}}\right),
    \end{align*}
    where $k^{\prime}=\min\{k,L\}$.
\end{lemma}

Lemma~\ref{sphere-packing-bound} is the binary version of the one in~\cite[Proposition 1]{rosenbloom1997codes} and in~\cite[Lemma 3]{jain2013}, and  since it was proved there, we omit the proof.
A simplified expression for the case where $e\le L$ is given next.
\begin{lemma}\label{small-e-sphere-packing-bound}
    The volume of a ball of radius $r$ in $\mathbb{F}_2^{n\times L}$, where $r\le L$ is,
    \begin{align*}
            V_{\text{TE}}(r) = 1 +\sum_{k=1}^{r}\sum_{i=1}^{k}\binom{n}{i}\binom{k-1}{i-1} 2^{k-i}~.
    \end{align*}
    And $V_{\text{TE}}(0)$ is defined to be equal to $1$.
\end{lemma}
\begin{IEEEproof}
Let $\bm{X}\in \mathbb{F}_2^{n\times L}$ and $k \leq n$ be the number of TEs that occurred in $\bm{X}$.
Moreover, let $i \leq k$ be the number of rows that suffer from TEs in $\bm{X}$.
Thus, there are $\binom{n}{i}$ possible combinations of rows that suffer from those $k$ TEs, and for each such possible combination of rows, there are $\binom{k-1}{i-1}$ ways to distribute $k$ erasures into $i$ rows.
Also, for each erroneous row, the first (leftmost) erased bit is known (must differ from $\bm{X}$), but the other bits afterwards and until the end of the row, can be either $1$ or $0$.
To conclude, for each $k$ and $i$, there are $\binom{n}{i}\binom{k-1}{i-1} 2^{k-i}$ possible distinct $\bm{Y}\in \mathbb{F}_2^{n\times L}$ such that $\rho_{\text{TE}}(\bm{X},\bm{Y})=k$ and there are exactly $i$ distinct rows between $\bm{X}$ and $\bm{Y}$.
Then, we complete the proof, using the fact that the number of arrays $\bm{Y}\in \mathbb{F}_2^{n\times L}$ such that $\rho_{\text{TE}}(\bm{X},\bm{Y})=k$ is
$$\sum_{i=1}^{k}\binom{n}{i}\binom{k-1}{i-1} 2^{k-i}~.$$

\end{IEEEproof}

Equipped with the previous lemma, we can obtain the following sphere packing bound.
\begin{theorem}\label{thm:TE-sphere-packing}
Let $\cC$ be a $(n\times L,M,d)_{\text{TE}}$ code. Then,
\begin{align*}
    M\cdot V_{\text{TE}}\left(\left\lfloor (d-1)/2\right\rfloor\right) \le 2^{n\cdot L}~.
\end{align*}
\end{theorem}
\noindent Next are some examples for the size of the above volume.
\begin{example}\label{ex:TE-spheres}
    The size of $V_{\text{TE}}(r)$, for $r=1,2,3$:
    \begin{align*}
        &V_{\text{TE}}(1) = 1+n,\\
        &V_{\text{TE}}(2) = 1+\frac{n^2+5n}{2},\\
        &V_{\text{TE}}(3) = 1+\frac{n^3+12n^2+29n}{6}.
    \end{align*}
    The results above are used to derive the bound from Theorem~\ref{thm:TE-sphere-packing} for $d=3,5,7$, and the authors achieve a tighter bound than the one in Claim \ref{clm:first-TE-bound}, as can be verified easily.
\end{example}

In Table~\ref{Summary-TE-Table}, one can find a summary of Corollary~\ref{cor:TE-redundancy-upper-bound}, Claims~\ref{clm:TE-3-4} and~\ref{clm:TE-5}, Claim~\ref{clm:first-TE-bound} and Theorem~\ref{thm:TE-sphere-packing}.

Finally, we note for completeness that in~\cite{rosenbloom1997codes}, a version of the Singleton bound was derived for the $m$-metric, although it is mostly relevant to cases of larger alphabets. 
The result is as follows.
\begin{claim}\label{clm:Singleton-bound}
    Let $\cC$ be a $(n\times L,M,d)_{\text{TE}}$ code. Then, 
    $$d \le nL - \lceil\log_2(M)\rceil + 1~.$$
\end{claim}

\subsection{Upper Bounds on the Cardinality of \texorpdfstring{$(t,s)$}{(t,s)}-DC Codes}
\label{sec:Bounds:DC Upper Bounds}
In the following subsection, three upper bounds on the cardinality of $(t,s)$-DC codes are derived.
First, we give an overview of the results, in the order it will be proved afterwards through this section.
Moreover, a summary of the bounds for $(t,s)$-DC codes is in Table~\ref{Summary-DC-Table}.
This includes both the constructive upper bounds from Section~\ref{sec:(t,s)-del-corr} and the lower bounds that are derived next.

Let $\cA_{\text{DC}}(t,s)$ be the largest cardinality of a $(t,s)$-DC code of dimension $n\times L$.
Furthermore, denote the largest cardinality of an $s$-deletion-correcting code of vectors of length $L$ over $\F_2$ by $M_s(L)$.\vspace{7pt}

\begin{theorem}\label{thm:summary-of-DC-bounds} 
    The following are bounds on $\cA_{\text{DC}}(t,s)$.
    \begin{enumerate}
        \item For any positive integers $s,t$,
        $$\cA_{\text{DC}}(t,s) \leq (M_s(L))^{t}\cdot 2^{L(n-t)}~.$$

        \item For integers $t,s \geq 2$,
        $$\cA_{\text{DC}}(t,s) \leq \frac{2^{nL}}{\left(\frac{n}{\lfloor t/2\rfloor}\cdot{\left(\frac{L}{\lfloor s/2\rfloor}\right)}^{\lfloor s/2\rfloor}\right)^{\lfloor t/2\rfloor}}~.$$

        \item For an integer $t \geq 2$,
        $$\cA_{\text{DC}}(t,1) \leq \frac{2^{nL}+1}{\left(\frac{n}{\lfloor t/2\rfloor}\right)^{\lfloor t/2\rfloor}}~.$$
    \end{enumerate}
\end{theorem}

\vspace{8pt}The bounds in Theorem~\ref{thm:summary-of-DC-bounds} imply lower bounds on the number of redundancy bits, as summarized next.
\begin{corollary}\label{cor:DC-red-lower-bounds}
    The asymptotic lower bounds on the redundancy of a $(t,s)$-DC code are as follows.
    \begin{enumerate}
        \item $t(L-\log_2(M_s(L))$.
        \item $\lfloor t/2 \rfloor (\log_2(n)+\lfloor s/2 \rfloor\log_2(L)) + O_{s,t}(1)$.
        \item $\lfloor t/2 \rfloor \log_2(n) + O_{s,t}(1)$, for $s=1$.
    \end{enumerate}
\end{corollary}
Note that by observing the case of $s=1$, and comparing to the results in Section~\ref{subsec:(t,1)-DC-const}, one can verify the following.
First, in~\cite{Fazeli-Vardy-Yaakobi-2015} a bound is presented, such that $M_1(L)\leq \frac{2^{L}-2}{L-1}$, and therefore a lower bound on the redundancy of a $(t,1)$-DC code is at least $t\log_2(L-1)$ bits, which is attained (up to one bit) in Construction~\ref{const:(t,1)-implicit-DC}, using an MDS code, when $n\leq 2^{h}+1$.
Secondly, when $n=c\cdot 2^{h}$ for some positive integer $c$, the constructive bound from Construction~\ref{const:(t,1)-implicit-DC} is $t\log_2(n)$, which, compared to the third lower bound in Corollary~\ref{cor:DC-red-lower-bounds}, differ only by a factor of 2.

Next, proofs for all the three bounds will be given, starting with the first one. However, before the proofs, we give the definition of the \emph{Fixed Length Levenshtein} (FLL) distance, as defined in~\cite{DBL21}, and repeated next. 
The FLL distance between two binary words of length $n$, denoted as $d_{\ell}$, is $s$, if a word $\boldsymbol{x}\in\F_2^{n}$ can be obtained from a word $\boldsymbol{y}\in\F_2^{n}$ using $s$ deletions and $s$ insertions (where $s$ is minimal).

\begin{IEEEproof}[Proof of Theorem~\ref{thm:summary-of-DC-bounds}, Part 1]
We want to show that $\cA_{\text{DC}}(t,s) \leq (M_s(L))^{t}\cdot 2^{L(n-t)}$.
Let $\cA^{\prime}_{\text{DC}}(t,s)$ denote the largest cardinality of a $(t,s)$-DC code, where the deletions can only occur in the first $t$ rows.
Since every $(t,s)$-DC code can also, in particular, correct $(t,s)$-DC errors when they occur specifically in the first $t$ rows, $\cA_{\text{DC}}(t,s) \leq \cA^{\prime}_{\text{DC}}(t,s)$, and thus providing an upper bound for $\cA^{\prime}_{\text{DC}}(t,s)$ is also an upper bound for $\cA_{\text{DC}}(t,s)$.

Next, let $\cC_{\text{DC}}$ be a $(t,s)$-DC code, where the deletions can only occur in the first $t$ rows, and of cardinality $\cA^{\prime}_{\text{DC}}(t,s)$.
We partition $\cC_{\text{DC}}$ into at most $2^{L(n-t)}$ subcodes, such that any two codewords arrays, $\bm{X},\bm{Y}\in\cC_{\text{DC}}$, that belong to the same subcode, have their last $n-t$ rows identical.
But, since $\bm{X},\bm{Y}$ belong to a code that corrects $(t,s)$-DC pattern in the first $t$ rows, $\bm{X}_{[1:t]}$ and $\bm{Y}_{[1:t]}$, the subarrays of $\bm{X}$ and $\bm{Y}$ that are obtained only from their first $t$ rows, belong to a $(t,s)$-DC code over $\F_2^{t\times L}$.
Thus, the cardinality of each subcode is at most $D(t,s,L)$, where $D(t,s,L)$ is the largest cardinality of a $(t,s)$-DC code over $\F_2^{t\times L}$, and there are $2^{L(n-t)}$ subcodes, which imply $\cA_{\text{DC}}(t,s) \leq \cA^{\prime}_{\text{DC}}(t,s) \leq D(t,s,L)\cdot 2^{L(n-t)}$, and we wish to prove that $D(t,s,L)\leq (M_s(L))^{t}$.

We provide a proof by induction on $t$.
First, for $t=1$, the code is over vectors of length $L$, and therefore the inequality $D(1,s,L) \leq M_s(L)$ is given by definition.
Next, we prove that if $D(i,s,L)\leq (M_s(L))^{i}$, then $D(i+1,s,L)\leq (M_s(L))^{i+1}$.
Let $\cC$ be an $(i+1,s)$-DC code over $\F_2^{(i+1)\times L}$ of cardinality $D(i+1,s,L)$, and let $\bm{X}^{1}$ denote the set of all possible rows of the first row of any codeword array in $\cC$.
If $|\bm{X}^{1}| \leq M_s(L)$, then for every $\bm{x}^{1} \in \bm{X}^{1}$, denote by $\cC(\bm{x}^{1})\subset\cC$, the subcode of $\cC$, in which the first row of every codeword array in $\cC(\bm{x}^{1})$ is $\bm{x}^{1}$.
Note that $\cC(\bm{x}^{1})$ can correct $s$ deletions in each of its $i+1$ rows, as a subcode of $\cC$, and since for every codeword in $\cC(\bm{x}^{1})$, its first row remains the same, it is necessary that the last $i$ rows of every codeword in $\cC(\bm{x}^{1})$ be an $(i,s)$-DC correcting code, or else there exists a pattern of deletions that one cannot recover from.

Therefore, $|\cC(\bm{x}^{1})|\leq D(i,s,L)$, which by the induction step, yields $|\cC(\bm{x}^{1})| \leq (M_s(L))^{i}$, and since $|\bm{X}^{1}| \leq M_s(L)$, we get that $D(i+1,s,L) = |\cC| \leq (M_s(L))^{i+1}$.

Next, if $|\bm{X}^{1}| > M_s(L)$, we will construct, in a greedy manner, another $(i+1,s)$-DC code, denoted as $\overline{\cC}$, where $|\overline{\cC}|=|\cC|$, and its set of all possible rows of the first row of any codeword array in the code $\overline{\cC}$, denoted $\overline{{\bm{X}}}^{1}$, is of cardinality at most $M_s(L)$, which then will finish the proof. 
First, arbitrarily choose any $\bm{y}_1 \in \overline{{\bm{X}}}^{1}$, and define the set 
$Y_{\bm{y}_1}=\left\{\bm{y} \in \overline{{\bm{X}}}^{1} \colon d_{\ell}(\bm{y},\bm{y}_1) \leq s\right\}$,
i.e., all the row vectors in $\overline{{\bm{X}}}^{1}$, that can be reached from $\bm{y}_1$ after at most $s$ deletions and insertions. Note that the set $Y_{\bm{y}_1}$ contains $\bm{y}_1$. Next, we select any vector $\bm{y}_2 \in \overline{{\bm{X}}}^{1} \setminus Y_{\bm{y}_1}$, and we define the set 
$Y_{\bm{y}_2}=\left\{\bm{y} \in \overline{{\bm{X}}}^{1} \setminus Y_{\bm{y}_1} \colon d_{\ell}(\bm{y},\bm{y}_2) \leq s\right\}$.

We keep repeating this procedure until we have $\nu$ sets $Y_{\bm{y}_1}$, $Y_{\bm{y}_2}$, $\ldots,$ $Y_{\bm{y}_\nu}$ (which are  disjoint by design) and where $Y_{\bm{y}_1} \cup \cdots \cup Y_{\bm{y}_\nu} = \overline{{\bm{X}}}^{1}$. By design, $\{\bm{y}_i\}_{i=1}^{\nu}$ is an $s$-deletion-correcting code and so $\nu \leq M_s(L)$.

Let $\cC(\{\bm{y}_{1},\bm{y}_{2},\ldots,\bm{y}_{k}\})$ be the subcode of $\cC$ where the first row of every codeword array in the subcode $\cC(\{\bm{y}_{1},\bm{y}_{2},\ldots,\bm{y}_{k}\})$ belongs to the set $\{\bm{y}_{1},\bm{y}_{2},\ldots,\bm{y}_{k}\}$.
Then, let $\overline{\cC}$ be defined in the following way. 
For every $i\in[\nu]$, we take all the codewords in $\cC(Y_{\bm{y}_{i}})$,  add those to $\overline{\cC}$, and set their first row as $\bm{y}_i$. 

We claim that $|\overline{\cC}|=|\cC|$ and that $\overline{\cC}$ is an $(i+1,s)$-DC code. 
\begin{enumerate}
    \item Assume, for the sake of contradiction, that $\overline{\cC}$ is not an $(i+1,s)$-DC code. Then, there exists $\overline{\bm{X}},\overline{\bm{Y}} \in\overline{\cC}$ such that these arrays are confusable after a pattern of $(i+1,s)$ deletions.
    Specifically, it means that their first rows, $\overline{\bm{x}}_1, \overline{\bm{y}}_1$ are confusable after $s$ deletions, and thus $\overline{\bm{x}}_1 = \overline{\bm{y}}_1$ as we chose the first row vectors from an $s$-deletion-correcting code.
    Hence, in $\cC$, there exist $\bm{X},\bm{Y}$ such that $\bm{X}$ and $\overline{\bm{X}}$ coincide on the last $i$ rows, and the same is true for $\bm{Y}$ and $\overline{\bm{Y}}$. 
    
    However, since $\overline{\bm{x}}_1 = \overline{\bm{y}}_1$, the first row of $\bm{X}$ and the first row of $\bm{Y}$ belong to the same set $Y_i$, which implies that $\bm{x}_1, \bm{y}_1$ are confusable after $s$ deletions, and since $\cC$ is an $(i+1,s)$-DC code, the last $i$ rows of $\bm{X}$ and $\bm{Y}$ cannot be confusable after a pattern of $(i,s)$ deletions, which also implies that $\overline{\bm{X}}$ and $\overline{\bm{Y}}$ cannot be confusable, as they have the same last $i$ rows.
    In turn, this means that $\overline{\bm{X}}$ and $\overline{\bm{Y}}$ are not confusable after $(i+1,s)$ deletions, and therefore a contradiction.

    \item We show that $|\overline{\cC}|=|\cC|$. 
    Since every codeword of $\overline{\cC}$ was constructed from a unique codeword in $\cC$, it implies a surjective function from $\cC$ to $\overline{\cC}$ and thus $|\overline{\cC}| \leq |\cC|$.
    Next, suppose, for the sake of contradiction, that $|\overline{\cC}|<|\cC|$, i.e., the function is not injective. 
    
    Then, we will have two codewords in $\cC$ that differ only in their first row (which was the only row that was changed), where one is $\bm{u}\in Y_i$ and the other is $\bm{v}\in Y_i$. 
    In this case, a pattern of $s$-deletions in the first row results two confusable arrays, a contradiction to $\cC$ being an $(i+1,s)$-DC code.
\end{enumerate}

To conclude, we constructed $\overline{\cC}$, an $(i+1,s)$-DC code, where $|\overline{\cC}|=|\cC|$, and its set of possible first rows is of cardinality at most $M_s(L)$.
Thus, $|\overline{{\bm{X}}}^{1}| \leq M_s(L)$ and $D(i+1,s,L) = |\cC| \leq (M_s(L))^{i+1}$.

\end{IEEEproof}

Next, a proof for the second bound is given. Since the second bound uses a sphere packing argument, a definition of a distance function is defined first. To do that, we will use the FLL distance.
\begin{definition}\label{def:d-DC}
    Let $\boldsymbol{X},\boldsymbol{Y}\in\F_{2}^{n\times L}$ and $s \geq 0$ is an integer. Then,
    \begin{align*}
        d_{s\text{-DC}}(\boldsymbol{X},\boldsymbol{Y})=
        \begin{cases}
            \infty & \text{if } \exists i\!\in\![n]\colon d_{\ell}(\boldsymbol{x}_i,\boldsymbol{y}_i)\!>\!s, \\
            |\{i\colon \boldsymbol{x}_i\ne\boldsymbol{y}_i\}| & \text{otherwise}.
        \end{cases}
    \end{align*}
\end{definition}

\begin{remark}\label{remark:d_DC-for-s=0}
If $s=0$, then for any $\boldsymbol{X},\boldsymbol{Y}\in\F_{2}^{n\times L}$, 
$$
    d_{0\text{-DC}}(\boldsymbol{X},\boldsymbol{Y})=
    \begin{cases}
        \infty & \text{if } \boldsymbol{X} \neq \boldsymbol{Y}~, \\
        0 & \text{otherwise}.
    \end{cases}
$$
\end{remark}

The following theorem states the important connection between Definition~\ref{def:d-DC} and $(t,s)$-DC codes.

\begin{theorem}\label{thm:DC-dist-is-suitable-for-DC-codes}
A code $\cC_{s\text{-DC}}\subseteq\F_{2}^{n\times L}$ is a $(t,s)$-DC code if and only if every $\boldsymbol{X},\boldsymbol{Y}\in\cC_{s\text{-DC}}$ satisfy $d_{s\text{-DC}}(\boldsymbol{X},\boldsymbol{Y})>t$.
\end{theorem}
\begin{IEEEproof}
First, let $\boldsymbol{X},\boldsymbol{Y}\in\cC_{s\text{-DC}}$ and assume for the sake of contradiction that $d_{s\text{-DC}}(\boldsymbol{X},\boldsymbol{Y}) \leq t$.
This implies that there are only $w\leq t$ indices of rows, denoted as $\cI = \{i_1,\ldots,i_w\}$, such that $\boldsymbol{x}_{i_j} \neq \boldsymbol{y}_{i_j}$ for every $i_j\in\cI$, and $\boldsymbol{x}_{k} = \boldsymbol{y}_{k}$ for every $k\notin\cI$.
Moreover, it also implies that $1 \leq d_{\ell}(\boldsymbol{x}_{i_j},\boldsymbol{y}_{i_j}) \leq s$ for every $i_j\in\cI$.
Thus, for every row index $i_j\in\cI$, one can delete $s$ bits in $\boldsymbol{x}_{i_j}$ and $s$ bits in $\boldsymbol{y}_{i_j}$ such that the resulted $(L-s)$-length rows are indistinguishable, as was stated in~\cite{DBL21}.
Since $|\cI|\le t$, there exists a pattern of $(t,s)$ deletions (i.e., choosing $t$ rows and then choose $s$ bits to delete in each of them) on $\boldsymbol{X}$ and another pattern of $(t,s)$ deletions (over the same row indices) on $\boldsymbol{Y}$, such that the resulted erroneous (punctured) arrays are indistinguishable, which is a contradiction for $\boldsymbol{X},\boldsymbol{Y}$ being both codewords of $\cC_{s\text{-DC}}$.

Conversely, let $\cC_{s\text{-DC}}$ be a code where every $\boldsymbol{X},\boldsymbol{Y}\in\cC_{s\text{-DC}}$ satisfy $d_{s\text{-DC}}(\boldsymbol{X},\boldsymbol{Y})>t$, and we will show that it is a $(t,s)$-DC code.
Assume, for the sake of contradiction, that there exist two patterns of $(t,s)$ deletions, $P_1$ and $P_2$, such that $\boldsymbol{X}$ suffers from deletions according to $P_1$ and $\boldsymbol{Y}$ suffers from deletions according to $P_2$, but the resulted erroneous arrays are indistinguishable.
Then, both $P_1$ and $P_2$ caused deletions in the same $w\le t$ rows, so we denote the indices of those rows as $\cI = \{i_1,\ldots,i_w\}$.
First, note that $\boldsymbol{x}_{i_j} = \boldsymbol{y}_{i_j}$ for every $i_j\notin\cI$.
Also, since every $i_j$-th row, where $i_j\in\cI$, suffer at most $s$ deletions, it implies $d_{\ell}(\boldsymbol{x}_{i_j},\boldsymbol{y}_{i_j}) \leq s$, otherwise the resulted erroneous arrays cannot be indistinguishable.
But, this means that $\forall i\in[n] \colon d_{\ell}(\boldsymbol{x}_i,\boldsymbol{y}_i) \leq s$ (either the rows are the same, or they differ by at most $s$ deletions), and there are at most $t$ rows such that $d_{\ell}(\boldsymbol{x}_i,\boldsymbol{y}_i) \ge s$, which is a contradiction, since $d_{s\text{-DC}}(\boldsymbol{X},\boldsymbol{Y})>t$.

\end{IEEEproof}

Following is the definition for the volume of a ball, using the distance function $d_{s\text{-DC}}$.

\begin{definition}\label{def:volume-dc}
    Let $\boldsymbol{X}\in\F_{2}^{n\times L}$. Then, the volume of an $(\tau,\sigma)$-ball, centered at $\boldsymbol{X}$, using the distance function $d_{s\text{-DC}}$, is as follows,
\begin{align*}
    \text{V}_{\text{DC}}(\tau,\sigma,\boldsymbol{X}) = \{\boldsymbol{Y}\in\F_{2}^{n\times L} \colon d_{\sigma\text{-DC}}(\boldsymbol{X},\boldsymbol{Y}) \leq \tau\}~.
\end{align*}
\end{definition}
A sphere packing argument follows immediately.

\begin{theorem}\label{thm:DC-sphere-packing}
    Let $\cC_{s\text{-DC}}=\{\boldsymbol{X}_i\}_{i=1}^{M}$ be a $(t,s)$-DC code of cardinality $M$. Then,
    \begin{align*}
        \sum_{i=1}^{M}\left|\text{V}_{\text{DC}} (\lfloor t/2\rfloor, \lfloor s/2\rfloor, \boldsymbol{X}_i)\right| \leq 2^{nL}~.
    \end{align*}
\end{theorem}
\begin{IEEEproof}
    It is sufficient to prove that the $(\lfloor t/2\rfloor,\lfloor s/2\rfloor)$-balls around each codeword, do not intersect.
    Assume, for the sake of contradiction, that there exists 
    $\boldsymbol{Y}\in\F_{2}^{n\times L}$, such that $\boldsymbol{Y} \in \text{V}_{\text{DC}}(\lfloor t/2\rfloor, \lfloor s/2\rfloor ,\boldsymbol{X}_i) \cap \text{V}_{\text{DC}}(\lfloor t/2\rfloor, \lfloor s/2\rfloor ,\boldsymbol{X}_j)$.
    Since $d_{\lfloor s/2\rfloor\text{-DC}}(\boldsymbol{Y},\boldsymbol{X}_i)<\infty$, there are at most $\lfloor t/2\rfloor$ distinct rows between $\boldsymbol{Y}$ and $\boldsymbol{X}_i$, and their FLL distance is at most $\lfloor s/2\rfloor$.
    Also, since $d_{s\text{-DC}}(\boldsymbol{Y},\boldsymbol{X}_j)<\infty$, there are at most $\lfloor t/2\rfloor$ distinct rows between $\boldsymbol{Y}$ and $\boldsymbol{X}_j$, and their FLL distance is at most $\lfloor s/2\rfloor$.
    
    Combining those two observations, there are at most $2\lfloor t/2\rfloor \leq t$ distinct rows between $\boldsymbol{X}_i$ and $\boldsymbol{X}_j$, and their FLL distance is at most $2\lfloor s/2\rfloor \leq s$.
    But this means that $d_{s\text{-DC}}(\boldsymbol{X}_i,\boldsymbol{X}_j)\le t$, which is a contradiction to Theorem~\ref{thm:DC-dist-is-suitable-for-DC-codes}.
    
\end{IEEEproof}

\noindent Since $\text{V}_{\text{DC}}(r_1,r_2,\boldsymbol{X})\!=\!\dot\bigcup_{i=0}^{r_1}\{\boldsymbol{Y}\!\colon d_{r_2\text{-DC}}(\boldsymbol{X},\boldsymbol{Y})\!=\!i\}$, the next claim will be a step towards estimating the volume of a ball.

\begin{claim}\label{claim:DC-sphere-size-t-dist-bound}
    Let $\boldsymbol{X}\in\F_{2}^{n\times L}$. Then, for any positive integer $s$,
    $|\{\boldsymbol{Y} \colon d_{s\text{-DC}}(\boldsymbol{X},\boldsymbol{Y}) = t\}| \geq \binom{n}{t}{\binom{L}{s}}^{t}$.
\end{claim}
\begin{IEEEproof}
    First, each $\boldsymbol{Y}$ such that $d_{s\text{-DC}}(\boldsymbol{X},\boldsymbol{Y}) = t$ is composed of $t$ row indices $\cI=\{i_{1},\ldots,i_{t}\}$ where $d_{\ell}(\boldsymbol{x}_i,\boldsymbol{y}_i) \le s$ and all the other row indices, $j\in [n] \setminus \cI$, satisfy $\boldsymbol{x}_i = \boldsymbol{y}_i$.
    Therefore, one can divide those binary arrays $\boldsymbol{Y}$ into $\binom{n}{t}$ distinct sets, that differ by their indices set $\cI$.
    Next, let $\cI$ be one of these sets of indices. 
    Then, the cardinality of $\cB_{\cI}(\boldsymbol{X})=\{\boldsymbol{Y} \mid \forall i\in\cI \colon d_{\ell}(\boldsymbol{x}_i,\boldsymbol{y}_i) \le s, \text{ and } \forall i\notin\cI \colon \boldsymbol{x}_i = \boldsymbol{y}_i\}$ is $\prod_{i\in\cI}|\cL_s(\boldsymbol{x}_i)|$, where $\cL_s(\boldsymbol{x}_i) = \{\boldsymbol{y}_i \colon d_{\ell}(\boldsymbol{x}_i,\boldsymbol{y}_i) \le s\}$.
    Using~\cite[Corollary 3]{DBL21}, $|\cL_s(\boldsymbol{x}_i)| \geq \binom{L}{s}$ and thus $\prod_{i\in\cI}|\cL_s(\boldsymbol{x}_i)| \ge {\binom{L}{s}}^{t}$.
    Since there are $\binom{n}{t}$ distinct sets, we conclude the proof.
    
\end{IEEEproof}

\begin{remark}
    It is clear from Remark~\ref{remark:d_DC-for-s=0} that for any positive integer $t$, $|\{\boldsymbol{Y} \colon d_{0\text{-DC}}(\boldsymbol{X},\boldsymbol{Y}) = t\}| = 0$,  and thus $|\text{V}_{\text{DC}}(r,0,\boldsymbol{X})| = 1$ for any integer $r$.
\end{remark}

Next, a lemma is given, to conclude a lower bound for the volume of a ball.

\begin{lemma}\label{lemma:lower-bound-on-DC-sphere}
Let $\boldsymbol{X}\in\F_{2}^{n\times L}$. Then, for any integer $s \ge 2$.
\begin{align*}
    |\text{V}_{\text{DC}} (\lfloor t/2\rfloor, \lfloor s/2\rfloor, \boldsymbol{X})| \ge \sum_{i=0}^{\lfloor t/2\rfloor}\binom{n}{i}{\binom{L}{\lfloor s/2\rfloor}}^{i}~.
\end{align*}
\end{lemma}
\begin{IEEEproof}
Using Definition~\ref{def:volume-dc} and Claim~\ref{claim:DC-sphere-size-t-dist-bound},
    \begin{align*}
    |\text{V}_{\text{DC}} (\lfloor t/2\rfloor, \lfloor s/2\rfloor, \boldsymbol{X})|\!&=\!\sum_{i=0}^{\lfloor t/2\rfloor}  |\{\boldsymbol{Y}\!\colon\!d_{(\lfloor s/2\rfloor)\text{-DC}}(\boldsymbol{X},\boldsymbol{Y})\!=\!i\}| \\
    &\ge \sum_{i=0}^{r}\binom{n}{i}{\binom{L}{\lfloor s/2\rfloor}}^{i}~.
\end{align*}

\end{IEEEproof}

Next, follows the proof of the second bound.
\begin{IEEEproof}[Proof of Theorem~\ref{thm:summary-of-DC-bounds}, Part 2]
    We show that if $\cC_{s\text{-DC}}$ is a $(t,s)$-DC code, $|\cC_{s\text{-DC}}|=M$, $s \geq 2$, then,
    \begin{align*}
        M\cdot \left(\frac{n}{\lfloor t/2\rfloor}\cdot{\left(\frac{L}{\lfloor s/2\rfloor}\right)}^{\lfloor s/2\rfloor}\right)^{\lfloor t/2\rfloor} \leq 2^{nL}~.
    \end{align*}
    Using Theorem~\ref{thm:DC-sphere-packing},
        \begin{align*}
            2^{nL} \geq \sum_{i=1}^{M}\left|\text{V}_{s\text{-DC}} \left(\left\lfloor\frac{t}{2}\right\rfloor,\boldsymbol{X}_i\right)\right|~.
        \end{align*}
    Then,  from Lemma~\ref{lemma:lower-bound-on-DC-sphere},
        \begin{align*}
            \sum_{i=1}^{M}\left|\text{V}_{s\text{-DC}}\left(\left\lfloor\frac{t}{2}\right\rfloor,\boldsymbol{X}_i\right)\right| \geq M\cdot\sum_{i=0}^{\lfloor t/2\rfloor}\binom{n}{i}{\binom{L}{\lfloor s/2\rfloor}}^{i}~.
        \end{align*}
    Then,
        \begin{align*}
         M\cdot\sum_{i=0}^{\lfloor t/2\rfloor}\binom{n}{i}{\binom{L}{\left\lfloor \frac{s}{2}\right\rfloor}}^{i}
         \geq M\binom{n}{\left\lfloor \frac{t}{2}\right\rfloor}{\binom{L}{\left\lfloor \frac{s}{2}\right\rfloor}}^{\left\lfloor \frac{t}{2}\right\rfloor}~.
        \end{align*}
    And to conclude,
    \begin{align*}
        M\binom{n}{\left\lfloor \frac{t}{2}\right\rfloor}{\binom{L}{\left\lfloor \frac{s}{2}\right\rfloor}}^{\left\lfloor \frac{t}{2}\right\rfloor}
        \geq M\left(\frac{n}{\left\lfloor \frac{t}{2}\right\rfloor}\!\cdot\!{\left(\frac{L}{\left\lfloor \frac{s}{2}\right\rfloor}\right)}^{\left\lfloor \frac{s}{2}\right\rfloor}\right)^{\left\lfloor \frac{t}{2}\right\rfloor}~.
    \end{align*}
\end{IEEEproof}

Finally, the third bound is derived. 
It is needed, in addition to the previous bound, since, as one can observe, the previous bound is not useful for the $(t,1)$-DC case.
Firstly, a definition of a simple distance function between vectors is given.
\begin{definition}
    Let $\boldsymbol{x},\boldsymbol{y}\in\F_{2}^{L}$. Then,
    \begin{align*}
        d^{1}(\boldsymbol{x},\boldsymbol{y})=
        \begin{cases}
            \infty & \text{if } \exists j\in\{2,\ldots,L\} \colon x_j \neq y_j~, \\
            1 & \text{if } x_1 \neq y_1\text{ and } \forall j\!\in\!\{2,\ldots,L\} \colon x_j = y_j ~, \\
            0 & \text{otherwise}~.
        \end{cases}
    \end{align*}
\end{definition}

Then, based on the above distance function, a definition of a distance function for arrays is given.

\begin{definition}
    Let $\boldsymbol{X},\boldsymbol{Y}\in\F_{2}^{n\times L}$. Then,
    \footnotesize\begin{align*}
        d^{1}_{\text{DC}}(\boldsymbol{X},\boldsymbol{Y})=
        \begin{cases}
            \infty & \text{if } \exists i\in[n]\colon d^{1}(\boldsymbol{x}_i,\boldsymbol{y}_i)=\infty~, \\
            |\{i \colon d^{1}(\boldsymbol{x}_i,\boldsymbol{y}_i)=1\}| & \text{otherwise}~.
        \end{cases}
    \end{align*}
\end{definition}

\normalsize
The next theorem demonstrates the relationship between the distance function $d^{1}_{\text{DC}}$ and $(t,1)$-DC codes.

\begin{theorem}\label{thm:d-prime-dist-is-suitable-for-DC-codes-of-s=1}
Let $\cC_{\text{DC}}\subseteq\F_{2}^{n\times L}$ be a $(t,1)$-DC code. Then, every $\boldsymbol{X},\boldsymbol{Y}\in\cC_{\text{DC}}$ satisfy $d^{1}_{\text{DC}}(\boldsymbol{X},\boldsymbol{Y})>t$.
\end{theorem}
\begin{IEEEproof}
    Assume, for the sake of contradiction, that there exists $\boldsymbol{X},\boldsymbol{Y}\in\cC_{\text{DC}}$ such that $d^{1}_{\text{DC}}(\boldsymbol{X},\boldsymbol{Y}) \leq t$.
    Then, since the distance is finite, $\forall i\in[n]\colon d^{1}(\boldsymbol{x}_i,\boldsymbol{y}_i) \leq 1$.
    But since $d^{1}_{\text{DC}}(\boldsymbol{X},\boldsymbol{Y}) \leq t$, there are $w \leq t$ row indices, denoted as $\cI$, in which $d^{1}(\boldsymbol{x}_i,\boldsymbol{y}_i)=1$ for all $i\in\cI$.
    Moreover, all $\forall j\notin\cI \colon \boldsymbol{x}_j=\boldsymbol{y}_j$.
    Thus, a deletion in the first bit of all the $w$ rows, whose indices are in $\cI$, which is a valid pattern of a $(t,1)$-DC error, will yield the same punctured arrays from $\boldsymbol{X}$ and $\boldsymbol{Y}$, which is a contradiction for $\cC_{\text{DC}}$ being a $(t,1)$-DC code.
    
\end{IEEEproof}

Following, as in the previous bound, is a sphere packing argument, starting with a definition of a ball.

\begin{definition}\label{def:volume-dc-prime}
    Let $\boldsymbol{X}\in\F_{2}^{n\times L}$. Then, the volume of a ball of radius $r$, centered at $\boldsymbol{X}$, using the distance function $d^{1}_{\text{DC}}$, is $\text{V}^{1}_{\text{DC}}(r,\boldsymbol{X}) = \{\boldsymbol{Y} \colon d^{1}_{\text{DC}}(\boldsymbol{X},\boldsymbol{Y}) \leq r\}.$
\end{definition}

Next, we prove the connection between balls in the Hamming distance and balls in the $d^{1}_{\text{DC}}$ distance.

\begin{claim}\label{clm:DC-prime-volume-argument}
Let $\boldsymbol{X}\in\F_{2}^{n\times L}$. Then, using $\text{V}_{2}(r,n)$ as defined in~\cite[Section 4.2]{RothBook2006}
\begin{align*}
    |\text{V}^{1}_{\text{DC}}(r,\boldsymbol{X})| = \text{V}_{2}(r,n) \delequal \sum_{i=0}^{r}\binom{n}{i}~.
\end{align*}
\end{claim}
\begin{IEEEproof}
    Since any two arrays $\boldsymbol{X},\boldsymbol{Y}\in\F_{2}^{n\times L}$ for which their $d^{1}_{\text{DC}}$ distance is finite, can differ only by bits in the first column, and then their $d^{1}_{\text{DC}}$ is equal to the Hamming distance of their first column, there is a clear bijection between the volume of balls in the Hamming distance, $\text{V}_{2}(r,n)$, and the volume of balls in the $d^{1}_{\text{DC}}$ distance function, $|\text{V}^{1}_{\text{DC}}(r,\boldsymbol{X})|$.
    
\end{IEEEproof}

Finally, the proof for the third bound follows.
\begin{IEEEproof}[Proof of Theorem~\ref{thm:summary-of-DC-bounds}, Part 3]
    We show that if $\cC_{\text{DC}}$ is a $(t,1)$-DC code of cardinality $M$, then,
    \begin{align*}
        M\cdot \text{V}_{2}(\lfloor t/2\rfloor,n)\leq 2^{nL}~.
    \end{align*}
    Again, it is sufficient to prove that the balls $\text{V}^{1}_{\text{DC}}(\lfloor t/2\rfloor,\cdot)$ around codewords of $\cC_{\text{DC}}$ do not intersect.
    Assume, for the sake of contradiction, that there exist an array $\boldsymbol{Y}\in \text{V}^{1}_{\text{DC}}(\lfloor t/2\rfloor,\boldsymbol{X}_1) \cap \text{V}^{1}_{\text{DC}}(\lfloor t/2\rfloor,\boldsymbol{X}_2)$, where $\boldsymbol{X}_1,\boldsymbol{X}_2\in\cC_{\text{DC}}$.
    Thus, $d^{1}_{\text{DC}}(\boldsymbol{X}_1,\boldsymbol{Y})<\infty$ and $d^{1}_{\text{DC}}(\boldsymbol{X}_1,\boldsymbol{Y})<\infty$, which implies that there are at most $\lfloor t/2\rfloor$ indices of rows where $\boldsymbol{X}_1$ and $\boldsymbol{Y}$ differ only in the first bit of the row, and the same is true for $\boldsymbol{X}_1$ and $\boldsymbol{Y}$.
    Therefore, there are at most $t$ rows in which $\boldsymbol{X}_1$ and $\boldsymbol{X}_2$ differ in their first bit, and all the other bits in those arrays are the same.
    So $d^{1}_{\text{DC}}(\boldsymbol{X}_1,\boldsymbol{X}_1)<t$, thus a contradiction to Theorem~\ref{thm:d-prime-dist-is-suitable-for-DC-codes-of-s=1}.
    
\end{IEEEproof}

\subsection{Upper Bounds on the Size of \texorpdfstring{$(1,1,1)$}{(1,1,1)}-TED Codes}
\label{sec:Bounds:TED Upper Bounds}
In order to investigate whether a code is optimal in terms of redundancy, a few notations are introduced. 

\begin{definition}
Let $\bm{X}\in\F_2^{n \times L}$ and denote $D^{\downarrow\uparrow}_i(\bm{X}) \subset \F_2^{n \times L}$ as the set of all possible arrays that can be received after deleting any bit from the $i$-th row of $\bm{X}$, and then inserting an arbitrary bit at the end of the same row of the resulting array.
\end{definition} 

Let $B_{\text{Del}}(\bm{x}_i)\subset\F_2^{L-1}$ be the $1$-deletion error ball of the $i$-th row of $\bm{X}$, i.e., the set of vectors that are obtained by removing one bit from $\bm{x}_i$. Moreover, let $(\bm{x}_i)^{\prime}\in B_{\text{Del}}(\bm{x}_i)$ be the vector that is obtained by one TE, i.e., $(\bm{x}_i)^{\prime}=(x_{i,1},\ldots,x_{i,L-1})$.
Therefore, for every $\bm{Z}\in D^{\downarrow\uparrow}_i(\bm{X})$, from the way $\bm{Z}$ is defined, $(\bm{z}_i)^{\prime} \in B_{\text{Del}}(\bm{x}_i)$. That is, $\bm{Z}\in D^{\downarrow\uparrow}_i(\bm{X})$ with the last element in the $i$-th row deleted, is a possible outcome of $\bm{X}$ after being transmitted through the $(1,1,1)$-TED channel.

\begin{example}
    Let 
    $$
        \bm{X} = \begin{bmatrix}
        0 & 1 & 1\\
        1 & 1 & 0\\
        0 & 0 & 1\\
        \end{bmatrix},~
        \bm{Y} = \begin{bmatrix}
        1 & 1 & 0\\
        0 & 1 & 1\\
        0 & 0 & 1\\
        \end{bmatrix}~.
    $$
    Then, 
    \footnotesize\begin{align*}
        D^{\uparrow \downarrow}_1(\bm{X})\!&=\!\left\{\!
        \begin{bmatrix}
        0 & 1 & 1\\
        1 & 1 & 0\\
        0 & 0 & 1\\
        \end{bmatrix}\!,\!
        \begin{bmatrix}
        0 & 1 & 0\\
        1 & 1 & 0\\
        0 & 0 & 1\\
        \end{bmatrix}\!,\!
        \begin{bmatrix}
        1 & 1 & 0\\
        1 & 1 & 0\\
        0 & 0 & 1\\
        \end{bmatrix}\!,\!
        \begin{bmatrix}
        1 & 1 & 1\\
        1 & 1 & 0\\
        0 & 0 & 1\\
        \end{bmatrix}\!\right\}\\
        D^{\uparrow \downarrow}_2(\bm{Y})\!&=\!\left\{\!
        \begin{bmatrix}
        1 & 1 & 0\\
        0 & 1 & 1\\
        0 & 0 & 1\\
        \end{bmatrix}\!,\!
        \begin{bmatrix}
        1 & 1 & 0\\
        0 & 1 & 0\\
        0 & 0 & 1\\
        \end{bmatrix}\!,\!
        \begin{bmatrix}
        1 & 1 & 0\\
        1 & 1 & 0\\
        0 & 0 & 1\\
        \end{bmatrix}\!,\!
        \begin{bmatrix}
        1 & 1 & 0\\
        1 & 1 & 1\\
        0 & 0 & 1\\
        \end{bmatrix}\!\right\}.
    \end{align*}
    \normalsize
    We want to show that $\bm{X}$ and $\bm{Y}$ cannot belong to the same $(1,1,1)$-TED code.
    Indeed, assume, for sake of contradiction, that they do.
    Let 
    $$
        \bm{Z} = \begin{bmatrix}
        1 & 1 & 0\\
        1 & 1 & 0\\
        0 & 0 & 1\\
        \end{bmatrix}
    \in D^{\uparrow \downarrow}_1(\bm{X}) \cap D^{\uparrow \downarrow}_2(\bm{Y})~.$$
    Therefore, let the array
    $$
        \widetilde{\bm{Z}} = \begin{bmatrix}
        1 & 1 & \bullet\\
        1 & 1 & \bullet\\
        0 & 0 & 1\\
        \end{bmatrix}~,
    $$
    be an output of the $(1,1,1)$-TED channel, i.e., $\bm{Z}$ with the first two rows suffer from a TE.
    Then, one cannot distinguish between the cases of a deletion in the first row of $\bm{X}$ and a TE in the second, and a TE in the first row of $\bm{Y}$ and a deletion in the second.
    Thus, $\bm{X}$ and $\bm{Y}$ cannot belong to the same $(1,1,1)$-TED code.
\end{example}

Next, let $\cD^{\uparrow \downarrow}(\bm{X}) = \bigcup_{i=1}^{n}D^{\uparrow \downarrow}_{i}(\bm{X})$ and the next claim follows, which
generalize the previous example.
\begin{claim}\label{clm:distinct-balls-TED}
Let $\cC \subseteq \F_2^{n \times L}$ be a $(1,1,1)$-TED code. Then, for any distinct $\bm{X},\bm{Y}\in\cC$,
\begin{align*}
\cD^{\uparrow \downarrow}(\bm{X}) \cap \cD^{\uparrow \downarrow}(\bm{Y}) = \emptyset~.
\end{align*}
\end{claim}
\begin{IEEEproof}
    We need to show that for any distinct $\bm{X}, \bm{Y}\!\in\!\cC$, we have $D^{\uparrow \downarrow}_{i_1}(\bm{X}) \cap D^{\uparrow \downarrow}_{i_2}(\bm{Y}) = \emptyset$, for any arbitrary row indices $i_1,i_2$.
    Assume, for sake of contradiction, that there exists an array $\bm{Z}\in D^{\uparrow \downarrow}_{i_1}(\bm{X}) \cap D^{\uparrow \downarrow}_{i_2}(\bm{Y})$.
    If $i_1=i_2$, then it implies that for any other row than the $i_1$-th row, $\bm{X}$ and $\bm{Y}$ coincide, and that $B_{\text{Del}}(\bm{x}_i) \cap B_{\text{Del}}(\bm{y}_i) \neq \emptyset$, and therefore there is a deletion in the $i_1$-th row that cannot be recovered, while any deletion in the other rows cannot be recovered, and therefore the contradiction.
    Next, assume $i_1 \neq i_2$. 
    This implies again that $\bm{X}$ and $\bm{Y}$ coincide on any row other than $i_1,i_2$.
    Moreover, $\bm{z}_{i_1}=\bm{y}_{i_1}$ and $\bm{z}_{i_2}=\bm{x}_{i_2}$.
    Also, $(\bm{z}_{i_1})^{\prime} \in B_{\text{Del}}(\bm{x}_{i_1})$ and $(\bm{z}_{i_2})^{\prime} \in B_{\text{Del}}(\bm{y}_{i_2})$.
    Thus, the array
    
    $$\widetilde{\bm{Z}} = 
    \begin{bmatrix}
        z_{1,1} & z_{1,2} & \cdots & z_{1,L-1} & z_{1,L} \\
        \vdots & \vdots & \ddots & \vdots & \vdots \\
        z_{i_1,1} & z_{i_1,2} & \cdots & z_{i_1,L-1} & \bullet \\
        \vdots & \vdots & \ddots & \vdots & \vdots \\
        z_{i_2,1} & z_{i_2,2} & \cdots & z_{i_2,L-1} & \bullet \\
        \vdots & \vdots & \ddots & \vdots & \vdots \\
        z_{n,1} & z_{n,2} & \cdots & z_{n,L-1} & z_{n,L} \\
    \end{bmatrix}$$
    
    is indeed a valid output of the transmission of either $\bm{X}$ or $\bm{Y}$ through the $(1,1,1)$-TED channel, while $\bm{X}$ suffers from an arbitrary deletion in the $i_1$-th row and a TE in the $i_2$-th row, or $\bm{Y}$ suffers from a TE in the $i_1$-th row and an arbitrary deletion in the $i_2$-th row.
    Thus, a contradiction to the fact that both $\bm{X}$ and $\bm{Y}$ are codewords in $\cC$.
    
\end{IEEEproof}

Next, we use Claim~\ref{clm:distinct-balls-TED}, to prove the following upper bound on the cardinality of $(1,1,1)$-TED codes.
\begin{theorem}\label{thm:bound-TED}
Let $\cA_{\text{TED}}(n\times L)$ be the cardinality of the maximal size $(1,1,1)$-TED code of $n\times L$ binary arrays. Then,
\begin{equation}
    \cA_{\text{TED}}(n\times L)\lesssim \frac{2^{nL}}{nL}~\text{,\quad i.e.,~ }\lim_{n\to\infty}\frac{\cA_{\text{TED}}(n\times L)}{\frac{2^{nL}}{nL}} \le 1~.\nonumber
\end{equation}
\end{theorem}
\begin{IEEEproof}
First, following the definition in~\cite{DBL21} of a \emph{run} in a vector, denoted as $r(\bm{x}_i),~\bm{x}_i\in\F_2^L$, a generalization is presented next.
Let $\bm{X}=(\bm{x}_1,\dots,\bm{x}_n)\in\F_2^{n\times L}$, then \emph{the sum of runs} is defined as $R(\bm{X}) = \sum_{i=1}^{n}r(\bm{x}_i)$.
It is possible to show that $|D_{i}^{\uparrow\downarrow}(\bm{X})|=2r(\bm{x}_{i})$, and then $\left|\cD^{\uparrow \downarrow}(\bm{X})\right|=2R(\bm{X})$.
Next, assume that $\mathcal{C}$ is a $(1,1,1)$-TED code of size $\cA_{\text{TED}}(n\times L)$, and then define the following,
\begin{align*}
    \mathcal{C}_0= \left\{ \bm{X}\in\cC \colon R(\bm{X})\ge \frac{nL}{2} - \sqrt{nL\ln(nL)}\right\}
\end{align*}
And $\mathcal{C}_1=\mathcal{C}\setminus\mathcal{C}_0$. 
From Claim~\ref{clm:distinct-balls-TED}, every distinct codeword array $\bm{X}\in\mathcal{C}$ results a distinct $\cD(\bm{X})$, thus
\begin{align*}
    2^{nL} &\geq \sum_{\bm{X}\in\mathcal{C}}|\cD^{\uparrow \downarrow}(\bm{X})| \\
    &= \sum_{\bm{X}\in\mathcal{C}_0}|\cD^{\uparrow \downarrow}(\bm{X})| + \sum_{\bm{X}\in\mathcal{C}_1}|\cD^{\uparrow \downarrow}(\bm{X})| \\
    &\geq \sum_{\bm{X}\in\mathcal{C}_0}2R(\bm{X}) \\
    &\geq |\mathcal{C}_0|\cdot \left(nL - 2\sqrt{nL\ln(nL)}\right)~,
\end{align*}
and therefore, 
\begin{align*}
    |\mathcal{C}_0| &\le \frac{2^{nL}}{nL - 2\sqrt{nL\ln(nL)}} \approx \frac{2^{nL}}{nL}~.
\end{align*}

Where the last step above is due to the fact that $$\frac{2^{nL}}{nL - 2\sqrt{nL\ln(nL)}} / \frac{2^{nL}}{nL} \xrightarrow{n\to\infty}=1~.$$

Denoting $N=nL$ and $g(N)=\frac{N}{2} - \sqrt{N\ln(N)}$, we say
\begin{align*}
    |\mathcal{C}_1| &\le \sum_{r=n}^{g(N)-1}|\{\bm{X}\in\{0,1\}^{n\times L}\colon R(\bm{X})=r\}| \\
    &\le \sum_{r=1}^{g(N)-1}|\{\bm{X}\in\{0,1\}^{N}\colon r(\bm{X})=r\}| \\
    &= \sum_{r=1}^{g(N)-1}2\binom{N-1}{r-1}~.
\end{align*}
Where $r(\bm{X})$ is the number of runs in the vectorized array $\bm{X}$, that is appending every row to the end of its previous row.
Moreover, the second inequality is due to the fact that any vectorized array can either have the same number of runs (if every row ends with the same bit as the next row starts with), or have an larger number of runs.
Therefore, any array that is in one of the sets on the l.h.s (of the second inequality), has to be in one of the sets on the r.h.s.
Now, using the lemma in~\cite{Levenshtein_SPD66}, $|\mathcal{C}_1| = o\left(\frac{2^N}{N^2}\right)$.
To conclude, $\frac{|\mathcal{C}_1|}{\frac{2^{nL}}{nL}} \xrightarrow[]{n\to\infty}0$ and therefore,
\begin{equation}
    |\mathcal{C}_0| + |\mathcal{C}_1| \lesssim \frac{2^{nL}}{nL}~. \nonumber
\end{equation}
\end{IEEEproof}

That means that the redundancy of an $n\times L$ binary array code capable of correcting at most a single deletion and a single TE is asymptotically at least $\log_2(n) + \log_2(L)$ bits. 

\section{Conclusions and Future Work}\label{sec:conclusions}
Our research has shown promising results. 
For TE and DC codes we provided some optimal constructions, using some interesting techniques.
The new coding schemes developed in this work show that it is possible to correct many frequently occurring error patterns by manipulating the parity check matrices of existing linear codes, and using ideas based on tensor product codes.
With the ever-increasing demand for data storage, our findings contribute to the growing body of research into alternative data storage solutions, offering a new and innovative approach. Further research is necessary to fully assess the potential of this technique, but our results are a step towards a future where DNA storage is a reality.

\section*{Acknowledgments}
The authors would like to acknowledge Prof. Ron M. Roth and Prof. Yuval Cassuto for their helpful reviews and comments on the paper. We extend special thanks to Prof. Roth for suggesting the expression in Lemma~\ref{small-e-sphere-packing-bound} and its proof.
We are also grateful to the reviewers and editor for their insightful comments and suggestions, which have improved the quality of this work.

\bibliographystyle{IEEEtran}
\bibliography{refs}

\begin{thebibliography}{10}
\providecommand{\url}[1]{#1}
\csname url@samestyle\endcsname
\providecommand{\newblock}{\relax}
\providecommand{\bibinfo}[2]{#2}
\providecommand{\BIBentrySTDinterwordspacing}{\spaceskip=0pt\relax}
\providecommand{\BIBentryALTinterwordstretchfactor}{4}
\providecommand{\BIBentryALTinterwordspacing}{\spaceskip=\fontdimen2\font plus
\BIBentryALTinterwordstretchfactor\fontdimen3\font minus \fontdimen4\font\relax}
\providecommand{\BIBforeignlanguage}[2]{{%
\expandafter\ifx\csname l@#1\endcsname\relax
\typeout{** WARNING: IEEEtran.bst: No hyphenation pattern has been}%
\typeout{** loaded for the language `#1'. Using the pattern for}%
\typeout{** the default language instead.}%
\else
\language=\csname l@#1\endcsname
\fi
#2}}
\providecommand{\BIBdecl}{\relax}
\BIBdecl

\bibitem{DNA23}
``{DNA} data storage market size is expected to reach usd 1926.7 million by 2028,'' \url{https://www.prnewswire.com/news-releases/}, [Online; accessed January-2023].

\bibitem{CGK12}
G.~M. Church, Y.~Gao, and S.~Kosuri, ``Next-generation digital information storage in dna,'' \emph{Science}, vol. 337, no. 6102, pp. 1628--1628, 2012.

\bibitem{GBCDLS13}
N.~Goldman~et al., ``Towards practical, high-capacity, low-maintenance information storage in synthesized dna,'' \emph{Nature}, vol. 494, no. 1435, pp. 77--80, 2013.

\bibitem{Iridia19}
P.~Predki and M.~Cassidy, ``Systems and methods for writing and reading data stored in a polymer,'' U.S. Patent 2020/0224264 A1, Jul. 2020.

\bibitem{VK04}
B.~Vasic and E.~Kurtas, \emph{Coding and signal processing for magnetic recording systems}.\hskip 1em plus 0.5em minus 0.4em\relax CRC press, 2004.

\bibitem{CKZEPK19}
\BIBentryALTinterwordspacing
K.~Chen, J.~Kong, J.~Zhu, N.~Ermann, P.~Predki, and U.~F. Keyser, ``Digital data storage using dna nanostructures and solid-state nanopores,'' \emph{Nano Letters}, vol.~19, no.~2, pp. 1210--1215, 2019. [Online]. Available: \url{https://doi.org/10.1021/acs.nanolett.8b04715}
\BIBentrySTDinterwordspacing

\bibitem{PascalAshwin2007}
A.~Ganesan and P.~O. Vontobel, ``On the existence of universally decodable matrices,'' \emph{IEEE Transactions on Information Theory}, vol.~53, no.~7, pp. 2572--2575, 2007.

\bibitem{zhou2016bch}
W.~Zhou, S.~Lin, and K.~A. Abdel-Ghaffar, ``{BCH} codes for the rosenbloom--tsfasman metric,'' \emph{IEEE Transactions on Information Theory}, vol.~62, no.~12, pp. 6757--6767, 2016.

\bibitem{rosenbloom1997codes}
M.~Y. Rosenbloom and M.~A. Tsfasman, ``Codes for the m-metric,'' \emph{Problemy Peredachi Informatsii}, vol.~33, no.~1, pp. 55--63, 1997.

\bibitem{dougherty2002maximum}
S.~Dougherty and M.~Skriganov, ``Maximum distance separable codes in the $\rho$ metric over arbitrary alphabets,'' \emph{Journal of Algebraic Combinatorics}, vol.~16, pp. 71--81, 07 2002.

\bibitem{raviv2020hierarchical}
N.~Raviv, M.~Schwartz, R.~Cohen, and Y.~Cassuto, ``Hierarchical erasure correction of linear codes,'' \emph{Finite Fields and Their Applications}, vol.~68, p. 101743, 2020.

\bibitem{jain2013}
S.~Jain, ``Array codes in m-metric correcting independent and clustered errors simultaneously,'' \emph{World Applied Sciences Journal}, vol.~22, pp. 36--40, 01 2013.

\bibitem{LSWY19}
A.~Lenz, P.~H. Siegel, A.~Wachter-Zeh, and E.~Yaakobi, ``Anchor-based correction of substitutions in indexed sets,'' in \emph{IEEE International Symposium on Information Theory (ISIT)}, 2019, pp. 757--761.

\bibitem{IC17}
K.~Immink and K.~Cai, ``Design of capacity-approaching constrained codes for {DNA}-based storage systems,'' \emph{IEEE Communication Letters}, vol.~22, no.~2, pp. 224--227, 2018.

\bibitem{YKGM18}
S.~Yazdi, H.~Kiah, R.~Gabrys, and O.~Milenkovic, ``Mutually uncorrelated primers for {DNA}-based data storage,'' \emph{IEEE Transactions on Information Theory}, vol.~64, no.~9, pp. 6283--6296, 2018.

\bibitem{PJSF20}
W.~Press, J.~Jones, J.~Schaub, and I.~Finkelstein, ``Hedges error-correcting code for {DNA} storage corrects indels and allows sequence constraints,'' \emph{Proceedings of the National Academy of Sciences}, vol. 117, no.~31, pp. 18\,489--18\,496, 2020.

\bibitem{LSWY20}
A.~Lenz, P.~Siegel, and E.~Yaakobi, ``Coding over sets for {DNA} storage,'' \emph{IEEE Transactions on Information Theory}, vol.~66, no.~4, pp. 2331--2351, 2019.

\bibitem{Hengjia22}
H.~Wei and M.~Schwartz, ``Improved coding over sets for {DNA}-based data storage,'' \emph{IEEE Transactions on Information Theory}, vol.~68, no.~1, pp. 118--129, 2022.

\bibitem{Boruchovsky23}
A.~Boruchovsky, D.~Bar-Lev, and E.~Yaakobi, ``{DNA}-correcting codes: End-to-end correction in {DNA} storage systems,'' in \emph{2023 IEEE International Symposium on Information Theory (ISIT)}, 2023, pp. 579--584.

\bibitem{Sima21}
J.~Sima, N.~Raviv, and J.~Bruck, ``On coding over sliced information,'' \emph{IEEE Transactions on Information Theory}, vol.~67, no.~5, pp. 2793--2807, 2021.

\bibitem{LM09}
Z.~Liu and M.~Mitzenmacher, ``Codes for deletion and insertion channels with segmented errors,'' \emph{IEEE Transactions on Information Theory}, vol.~56, no.~1, pp. 224--232, 2009.

\bibitem{AVF17}
M.~Abroshan, R.~Venkataramanan, and A.~Fàbregas, ``Coding for segmented edit channels,'' \emph{IEEE Transactions on Information Theory}, vol.~64, no.~4, pp. 3086--30\,982, 2017.

\bibitem{MW67}
B.~Masnick and J.~Wolf, ``On linear unequal error protection codes,'' \emph{IEEE Transactions on Information Theory}, vol.~13, no.~4, p. 600–607, 1967.

\bibitem{ZX05}
Z.~Zhou and C.~Xu, ``An improved unequal error protection turbo codes,'' in \emph{Proceedings of 2005 International Conference on Wireless Communications, Networking and Mobile Computing}, Piscataway, 2005, pp. 284--287.

\bibitem{BK81}
I.~Boyafunov and G.~Katsman, ``Linear unequal error protection codes,'' \emph{IEEE Transactions on Information Theory}, vol.~27, no.~2, pp. 168--175, 1981.

\bibitem{lin2004error}
S.~Lin and D.~J. Costello, \emph{Error Control Coding}, 2nd~ed.\hskip 1em plus 0.5em minus 0.4em\relax Pearson, 2004.

\bibitem{RothBook2006}
R.~Roth, \emph{Introduction to Coding Theory}.\hskip 1em plus 0.5em minus 0.4em\relax USA: Cambridge University Press, 2006.

\bibitem{Hasse1936}
\BIBentryALTinterwordspacing
H.~Hasse, ``Theorie der höheren differentiale in einem algebraischen funktionenkörper mit vollkommenem konstantenkörper bei beliebiger charakteristik.'' \emph{Journal für die reine und angewandte Mathematik}, vol. 175, pp. 50--54, 1936. [Online]. Available: \url{http://eudml.org/doc/149955}
\BIBentrySTDinterwordspacing

\bibitem{VT1965}
R.~Varshamov and G.~Tenengolts, ``Codes which correct single asymmetric errors (in russian),'' \emph{Automatika i Telemkhanika}, vol. 161, no.~3, pp. 288--292, 1965.

\bibitem{Wolf06}
J.~K. Wolf, ``An introduction to tensor product codes and applications to digital storage systems,'' in \emph{2006 IEEE Information Theory Workshop - ITW '06 Chengdu}, 2006, pp. 6--10.

\bibitem{Levenshtein_SPD66}
V.~I. Levenshtein, ``Binary codes capable of correcting deletions, insertions and reversals.'' \emph{Soviet Physics Doklady}, vol.~10, no.~8, pp. 707--710, feb 1966, doklady Akademii Nauk SSSR, V163 No4 845-848 1965.

\bibitem{Abdel-Ghaffar98}
K.~Abdel-Ghaffar and H.~Ferreira, ``Systematic encoding of the varshamov-tenengol'ts codes and the constantin-rao codes,'' \emph{IEEE Transactions on Information Theory}, vol.~44, no.~1, pp. 340--345, 1998.

\bibitem{Sima2020}
J.~Sima, R.~Gabrys, and J.~Bruck, ``Optimal systematic t-deletion correcting codes,'' in \emph{2020 IEEE International Symposium on Information Theory (ISIT)}, 2020, pp. 769--774.

\bibitem{Fazeli-Vardy-Yaakobi-2015}
A.~Fazeli, A.~Vardy, and E.~Yaakobi, ``Generalized sphere packing bound,'' \emph{IEEE Transactions on Information Theory}, vol.~61, no.~5, pp. 2313--2334, 2015.

\bibitem{DBL21}
D.~Bar-Lev, T.~Etzion, and E.~Yaakobi, ``On levenshtein balls with radius one,'' in \emph{2021 IEEE International Symposium on Information Theory (ISIT)}, 2021, pp. 1979--1984.

\end{thebibliography}

\appendix\label{appendix}
\tecodeiff*
\begin{IEEEproof}
Let $\bm{X}\in\cC$. If $\cC$ is an $e$-TE code, it means that there is no other $\bm{Y}\in\cC$ such that $\bm{X}^{(\bm{p})} = \bm{Y}^{(\bm{p})}$ for any $\bm{p}\in P(e,L,n)$, where $\|\bm{p}\|_{1}=t$, and $t \le e$. 
Assume, for the sake of contradiction, that such $\bm{p}$ exists. It means that if $\bm{X}$ suffers from a TE pattern $\bm{p}$ where $\|\bm{p}\|_{1}=t$, and $t \le e$, it cannot be distinguished from $\bm{Y}$ that suffers from the same TE pattern $\bm{p}$.
Therefore, this TE cannot be corrected, which yields a contradiction to $\cC$ being a $e$-TE code. So, conclude that $\rho_{\text{TE}}(\cC)>e$.
Conversely, if $\rho_{\text{TE}}(\cC)\ge e+1$, let $\bm{X}\in\cC$ 
and assume an $e$-TE occurred in $\bm{X}$, and we want to show that the code can correct it.
Assume, for the sake of contradiction, that there exists a $\bm{Y}\in\cC$ and $\bm{p}\in P(e,n,L)$ such that $\bm{X}^{(\bm{p})} = \bm{Y}^{(\bm{p})}$ and $\|\bm{p}\|_{1}\le e$, and therefore the code cannot correct this $e$-TE, since it cannot distinguish between $\bm{X}$ and $\bm{Y}$.
But then $\rho_{\text{TE}}(\bm{X},\bm{Y})\le e$ and this contradicts the minimum $\rho_{\text{TE}}$-distance of the code.
We conclude that after at most $e$-TE occurred, $\bm{X}$ is the only codeword that could be the result of those $e$-TEs, and conclude that $\cC$ can correct up to $e$-TEs.

\end{IEEEproof}

\begin{IEEEbiographynophoto}{Boaz Moav}
received the B.Sc. degree in computer science and mathematics and the M.Sc. degree from the Computer Science Department, Technion-Israel Institute of Technology in in 2021 and 2023, respectively, where he is currently pursuing the Ph.D. degree with the Henry and Marilyn Taub Faculty of Computer Science. His advisor is Prof. E. Yaakobi. His research interests include coding theory and algorithms for DNA storage systems.
\end{IEEEbiographynophoto}
\begin{IEEEbiographynophoto}{Ryan Gabrys}
is a scientist jointly affiliated with the Qualcomm Institute and Naval Information Warfare Center San Diego. His research interests broadly lie in the areas of theoretical computer science and electrical engineering, including bioinformatics, combinatorics, coding theory, and signal processing. He is particularly interested in inter-disciplinary problems that span multiple areas such as biology, mathematics, and systems design.
Gabrys received a Ph.D. from the University of California Los Angeles in 2014; a Master of Engineering from the University of California San Diego in 2010; and a B.S. in Computer Science and Mathematics from the University of Illinois Urbana-Champaign in 2005.
\end{IEEEbiographynophoto}
\begin{IEEEbiographynophoto}{Eitan Yaakobi}
(S'07--M'12--SM'17) is an Associate Professor at the Computer Science Department at the Technion --- Israel Institute of Technology. He also holds a courtesy appointment in the Technion's Electrical and Computer Engineering (ECE) Department. He received the B.A. degrees in computer science and mathematics, and the M.Sc. degree in computer science from the Technion --- Israel Institute of Technology, Haifa, Israel, in 2005 and 2007, respectively, and the Ph.D. degree in electrical engineering from the University of California, San Diego, in 2011. Between 2011-2013, he was a postdoctoral researcher in the department of Electrical Engineering at the California Institute of Technology and at the Center for Memory and Recording Research at the University of California, San Diego. His research interests include information and coding theory with applications to non-volatile memories, associative memories, DNA storage, data storage and retrieval, and private information retrieval. He received the Marconi Society Young Scholar in 2009 and the Intel Ph.D. Fellowship in 2010-2011. Between 2020 and 2023, he served as an Associate Editor for Coding snd Decoding for the \textsc{IEEE Transactions on Information Theory}. Since 2016, he is affiliated with the Center for Memory and Recording Research at the University of California, San Diego, and between 2018--2022, he was affiliated with the Institute of Advanced Studies, Technical University of Munich, where he held a four-year Hans Fischer Fellowship, funded by the German Excellence Initiative and the EU 7th Framework Program. Between August 2023 and January 2024, he was a Visiting Associate Professor at the School of Physical and Mathematical Sciences at Nanyang Technological University. He is a recipient of several grants, including the ERC Consolidator Grant and the EIC Pathfinder Challenge. 
\end{IEEEbiographynophoto}

\end{document}